\documentclass{article}
\usepackage{emulateapj,pstricks,apjfonts}

\righthead{SARAZIN}
\lefthead{COSMIC RAY ELECTRONS IN CLUSTERS OF GALAXIES}
\slugcomment{The Astrophysical Journal, in press}
\submitted{The Astrophysical Journal, in press}

\begin{document}

\title{The Energy Spectrum of Primary Cosmic Ray Electrons
in Clusters of Galaxies and Inverse Compton Emission}

\author{Craig L. Sarazin}
\affil{Department of Astronomy, University of Virginia, \\
P.O. Box 3818, Charlottesville, VA 22903-0818; \\
cls7i@virginia.edu}

\begin{abstract}
Models for the evolution of the integrated energy spectrum of primary cosmic
ray electrons in clusters of galaxies have been calculated, including the
effects of losses due to inverse Compton (IC), 
synchrotron, and bremsstrahlung emission, and
Coulomb losses to the intracluster medium (ICM).
The combined time scale for these losses
reaches a maximum of $\sim$$3 \times 10^9$ yr for electrons with a
Lorentz factor $\gamma \sim 300$.
A variety of models for the time evolution of particle injection are
considered, including models in which the electrons are all produced
at a single epoch in the past, models with continuous particle
acceleration, and combinations of these.
Analytical solutions are given for a number of limiting cases.
Numerical solutions are given for more general cases.
Only clusters in which there has been a substantial injection of relativistic
electrons since $z \la 1$ will have any significant population of primary
cosmic ray electrons at present.
For models in which all of the electrons were injected in the past,
there is a high energy cutoff $\gamma_{max}$ to the present electron
distribution.
At low energies where Coulomb losses dominate, the electron distribution
function $N( \gamma )$ tends to a constant value, independent of $\gamma$.
On the other hand, if electrons are being accelerated at present, the
energy distribution at high and low energies approaches steady state.
If the electrons are injected with a power-law distribution, the steady
state distribution is one power steeper at high energies and one
power flatter at low energies.
In models with a large initial population of particles, but also with
a significant rate of current particle injection, the electron
distributions are a simple combination the behavior of the initial
population models and the steady injection models.
There is a steep drop in the electron population at $\gamma_{max}$,
but higher energy electrons are present at a rate determined by the
current rate of particle injection.
Increasing the ICM thermal gas density decreases the number of low energy
electrons ($\gamma \la 100$).
If the magnetic field is greater than generally expected,
($B \ga 3$ $\mu$G), synchrotron losses will reduce the
number of high energy electrons.
A significant population of electrons with $\gamma \sim 300$, associated
with the peak in the particle loss time, is a generic feature of the
models, as long as there has been significant particle injection since
$z \la 1$.

The IC and synchrotron emission from these models was calculated.
In models with steady particle injection with a power-law exponent $p$,
the IC spectra relax into a steady-state form.
At low energies, the spectrum is a power-law with 
$\alpha \approx -0.15$, while at high energies $\alpha \approx -1.15$.
These two power-laws meet at a knee at $\nu \sim 3 \times 10^{16}$ Hz.
In models with no current particle injection, the cutoff in the electron
distribution at high energies ($\gamma \ge \gamma_{max})$ results in
a rapid drop in the IC spectrum at high frequencies.
In models in which the current rate of particle injection provides a
small but significant fraction of the total electron energy,
the spectra show an extended hump at low frequencies ($\nu \la 10^{17}$ Hz),
with a rapid fall off above $\nu \sim 10^{16}$ Hz.
However, they also have an extended hard tail of emission at high
frequencies, which has a power-law spectrum with a spectral index of
$\alpha \approx -1.15$.

In the models, EUV and soft X-ray emission are nearly ubiquitous.
This emission is produced by electrons with $\gamma \sim 300$, which
have the longest loss times.
The spectra are predicted to drop rapidly in going from the EUV to
the X-ray band.
The IC emission also extends down the UV, optical, and IR bands,
with a fairly flat spectrum ($-0.6 \ga \alpha \ga +0.3$).
At hard X-ray energies $\ga$20 keV, IC emission should become observable 
against the background of thermal X-ray emission.
Such hard X-ray (HXR) emission is due to high energy electrons
($\gamma \sim 10^4$).
The same electrons will produce diffuse radio emission (cluster radio
halos) via synchrotron emission.
Because of the short loss times of these particles,
HXR and diffuse radio emission are only expected in clusters which have
current (or very recent) particle acceleration.
Assuming that the electrons are accelerated in ICM shocks, one would only
expect diffuse HXR/radio emission in clusters which are currently
undergoing a large merger.
The luminosity of HXR emission is primarily determined by the current
rate of particle acceleration in the cluster.
The spectra in most models with significant HXR or radio emission are
approximately power-laws with $\alpha \approx -1.1$.
The IC spectrum from the EUV to HXR are not generally fit by a single
power-law.
Instead, there is a rapid fall-off from EUV to X-ray energies, with
a power-law tail extending into the HXR band.
\end{abstract}


\section{Introduction} \label{sec:intro}

Early X-ray observations with the $Uhuru$ satellite
established that clusters of galaxies were luminous, extended
X-ray sources
(Cavaliere, Gursky, \& Tucker 1971). 
The X-ray emission is due to hot ($\sim$$10^8$ K), diffuse intracluster
gas
(see Sarazin [1988] for a review).
This intracluster medium (ICM) contains heavy elements with
abundances, relative to hydrogen, of about one third of the solar
values.
The gas is thought to have originated mainly from infall of intergalactic
gas, although the heavy elements probably came from gas ejected by
galaxies.
The gas is believed to have been shock-heated during infall into the
cluster, during subcluster mergers, and as a portion of it was ejected by
galaxies.

It is useful to compare the intracluster medium with the interstellar
medium which fills the volumes of space between stars in our Galaxy.
The interstellar medium in our Galaxy contains a mixture of at least
three different types of matter: thermal gases of various
densities and temperatures; solid dust grants;
and relativistic materials, including magnetic fields and
cosmic ray particles.
To what extent would one expect to find similar materials in the
intracluster medium?
It may be difficult for dust grains and cooler thermal gas phases to
survive in the hot ICM.
However, there is no obvious reason why clusters should not contain
magnetic fields and relativistic particles.

In fact, there are a number of reasons to think that cosmic ray particles
might be particularly abundant within the intracluster medium.
First, clusters of galaxies should be very effective traps for cosmic ray
ions and electrons.
Under reasonable assumptions for the diffusion coefficient, particles
with energies of less than $\la$$10^6$ GeV have diffusion times which
are longer than the Hubble time
(Berezinsky, Blasi, \& Ptuskin 1997;
Colafrancesco \& Blasi 1998).
Second, the lifetimes of cosmic ray particles, even the electrons
which are responsible for most of the radiative signatures of relativistic
particles, can be quite long.
The radiation fields (optical/IR and X-ray) and magnetic fields
($B \la 1 \, \mu$G) in the ICM are low enough that high energy electrons
mainly lose energy by inverse Compton (IC) scattering of Cosmic Microwave
Background (CMB) photons
(Sarazin \& Lieu 1998).
Lower energy electrons can lose energy by Coulomb interactions with
the plasma;
however, at the very low densities ($n_e \la 10^{-3}$ cm$^{-3}$) in the
bulk of the ICM, this is only important for electrons with
$ \gamma \la 200$
(Sarazin \& Lieu 1998).
(Here, $\gamma$ is the Lorentz factor for the electrons, so that their
total energy is $\gamma m_e c^2$.)
For electrons with $\gamma \ga 200$, the lifetime is set by IC losses and
is
\begin{equation} \label{eq:lifetime}
t_{IC} = \frac{(\gamma - 1 ) m_e c^2}{\frac{4}{3} \sigma_T c \gamma^2
U_{CMB}}
\approx
7.7 \times 10^{9}
\left( \frac{\gamma}{300} \right)^{-1} \,
{\rm yr}
\, .
\end{equation}
The lifetimes of ions are set by interactions; for protons, this gives
$t_{ion} \ga 10^{11} ( n_e / 10^{-3} \, {\rm cm}^{-3})^{-1}$ yr.
Thus, clusters of galaxies can retain low energy electrons
($\gamma \sim 300$) and nearly all cosmic ray ions for a significant
fraction of a Hubble time.

Cluster of galaxies are likely to have substantial sources of cosmic rays.
Clusters often contain powerful radio galaxies, which may
produce and distribute cosmic rays throughout the cluster, in addition
to possible contributions from the cluster galaxies.
However, the primary reason why the cosmic ray populations in clusters
might be large is connected with the high temperature of the intracluster
gas.
This indicates that all of the intracluster medium
(typically, $10^{14} \, M_\odot$ of gas) has passed through strong
shocks with shock velocities of $\sim$1000 km/s during its history.
In our own Galaxy, whenever diffuse gas undergoes a strong shock
at velocities of this order, a portion of the shock energy
goes into the acceleration of relativistic particles
(Blandford \& Ostriker 1978;
Bell 1978a,b;
Blandford \& Eichler 1987;
Jones \& Ellison 1991).
Thus, it seems likely that relatively efficient particle acceleration also
occurs in clusters of galaxies.

Direct evidence for the presence of an extensive population of
relativistic particles and magnetic fields in the ICM comes from
the observation of diffuse synchrotron radio halos in clusters
(e.g., Wilson 1970;
Hanisch 1982;
Giovannini et al.\ 1993;
Deiss et al.\ 1997).
More recently, extreme ultraviolet (EUV) and very soft X-ray emission
has been detected from a number of clusters
(Lieu et al.\ 1996a,b;
Mittaz, Lieu, \& Lockman 1998).
Although the origin and even the existence of this radiation remain
controversial,
one hypothesis is that it is inverse Compton (IC) emission by relativistic
electrons
(Hwang 1997;
En{\ss}lin \& Biermann 1998;
Sarazin \& Lieu 1998).
Finally, there are a number of reports of detections of hard
X-ray emission from clusters of galaxies with $BeppoSAX$, which
might be due to IC emission from higher energy electrons
(Fusco-Femiano et al.\ 1998;
Kaastra, Bleeker, \& Mewe 1998),
although this interpretation is still uncertain
(Henriksen 1998;
Goldoni et al.\ 1998).
The EUV emission would require electrons with $\gamma \sim 300$, while
the hard X-ray emission would require $\gamma \sim 10^4$.

The $BeppoSAX$ detection of hard X-ray emission from Coma implies that the
average magnetic field over large volumes of the cluster is 0.16 $\mu$G.
Faraday rotation measurements towards Coma and other clusters suggest
that the magnetic field in the core of the cluster is stronger,
perhaps around 1 $\mu$G
(Kim et al.\ 1990;
Kim, Kronberg, \& Tribble 1991;
Kronberg 1994).
In our models, we will use 0.3 $\mu$G as a typical value for the
intracluster magnetic field on large scales.

The energy spectra of relativistic particles in many astrophysical
environments are often modeled as power-laws.
The energy spectrum of cosmic ray electrons in clusters of galaxies
have been represented as power-laws in most investigations
(e.g,
Hwang 1997;
En{\ss}lin \& Biermann 1998).
This should be a reasonable approximation at energies where all of the
particles are relativistic and where there are no other physical scales
affecting the energy.
However, there are a number of reasons why the populations of relativistic
electrons in clusters of galaxies might not be well-represented by
power-laws.
First, at the low energies of interest for the EUV emission
($\gamma \sim 300$, or $E \sim 150$ MeV, where $E$ is the particle kinetic
energy), the electrons are relativistic, but the ions are not.
Second, at the energies interest for the EUV emission ($\gamma \sim 300$),
the life times of the electrons are close to the Hubble time
(eq.~\ref{eq:lifetime}).
At much higher or lower energies, the electron life times are
much shorter.
Thus, losses affect different parts of the electron energy spectrum in
different ways.
Finally, at low energies $\gamma \la 200$, the electrons mainly lose energy
to the plasma due to Coulomb losses.
At higher energies $\gamma \ga 200$, the most important losses are due
to IC or synchrotron emission.
These different loss mechanisms have different energy dependences, and
this breaks any simple scaling in the particle losses.

In this paper, illustrative models will be calculated for
the energy spectrum of primary, relativistic electrons.
Recently, models in which the relativistic electrons are secondaries
produced by interactions of cosmic ray ions have been given by
Colafrancesco \& Blasi (1998).
Cosmic ray ions and secondary electrons will be included in a subsequent
set of calculations.
In this paper, we will concentrate on the total energy spectrum
of all the relativistic electrons in the cluster, under the assumption
that they remain trapped in the cluster
(Berezinsky et al. 1997;
Colafrancesco \& Blasi 1998).
In a subsequent paper, we will discuss the spatial distribution
of the cosmic ray particles.
In this paper, radio synchrotron and inverse Compton emission spectra will
be determined up to the hard X-ray spectral band.
Gamma-ray emission by the same electrons (as well as emission by ions
through $\pi^o$ decay and other nuclear reactions) will be included in
a later paper.

\section{Evolution of the Electron Spectrum} \label{sec:espect}

\subsection{Evolution Equation} \label{sec:espect_eqn}

The evolution of the cosmic ray electron population in clusters is given
by the diffusion-loss equation (Ginzburg \& Syrovatskii 1964):
\begin{equation} \label{eq:diff_loss}
\frac{d \, n(E)}{d t} =
- n(E) \mbox{\boldmath $\nabla \cdot $} {\bf v}
+ \mbox{\boldmath $\nabla \cdot $}
\left[ D(E) \mbox{\boldmath $\nabla $} n(E) \right]
+ \frac{\partial}{\partial E} \left[ b(E)  n(E) \right]
+ q(E)
\, ,
\end{equation}
where $n ( E ) dE$ is the number density of electrons with kinetic energies in
the range $E$ to $E + dE$,
{\bf v} is the velocity of the ICM,
$d / d t$ is the Lagrangian time derivative,
$D (E) $ is the diffusion coefficient,
and $q(E)dE$ gives the rate of production of new cosmic ray particles
per unit volume with energies in the range $E$ to $E + dE$.
The losses by the particles are given by
$ b ( E ) \equiv - ( d E / d t)$, where the derivative gives the rate of
change in the energy of a single particle with an energy of $E$.
This formulation of the losses assumes that they are continuous;
this is a reasonable approximation for the Coulomb and inverse Compton
losses which are most important for electrons in clusters, but is not
completely correct for bremsstrahlung losses.
Equation~(\ref{eq:diff_loss}) ignores the acceleration of the electrons
(which we will treat separately and include in $q$)
and their generation as secondaries through the interaction of other
particles.

Here, we consider models for the total cosmic ray electron spectrum of
clusters, without considering the spatial distribution of the 
particles.
Let $N ( E ) dE$ be the total number of electrons in the cluster with
kinetic energies in the range $E$ to $E + dE$.
I will also assume that cosmic ray electrons and ions are trapped
within clusters, and cannot escape.
Models for cosmic ray diffusion within clusters do indeed suggest that
they should be trapped
(Berezinsky et al. 1997;
Colafrancesco \& Blasi 1998).
Then, integrating equation~(\ref{eq:diff_loss}) over the volume of the
cluster leads to an equation for the time evolution of the total electron
spectrum,
\begin{equation} \label{eq:evolution_energy}
\frac{\partial N(E)}{\partial t} =
\frac{\partial}{\partial E} \left[ b(E)  N(E) \right]
+Q(E)
\, ,
\end{equation}
where $Q(E) dE$ gives the total rate of production of new cosmic ray
electrons in the energy range $E$ to $E + dE$.
I have replaced $b(E)$ by its value averaged over all the particles in
the cluster.
The first two terms on the right hand side of equation~(\ref{eq:diff_loss})
are converted by Green's formula into fluxes at the outer surface of the
cluster, which are zero since we assume that the cluster retains all of
the particles.

I will use the Lorentz factor $\gamma$ of the electrons as the
independent variable rather than the kinetic energy
$E = ( \gamma - 1 ) m_e c^2$, where $m_e$ is the electron mass.
With the definitions
$N( \gamma ) d \gamma = N(E) dE$,
$Q( \gamma ) d \gamma = Q(E) dE$,
and
$b( \gamma ) = b( E ) / m_e c^2$,
the expression for the energy loss by an individual particle becomes
\begin{equation} \label{eq:loss}
\frac{ d \, \gamma}{d \, t} = - b ( \gamma , t ) \, ,
\end{equation}
and the equation for the evolution of the electron populations is
\begin{equation} \label{eq:evolution}
\frac{\partial N( \gamma )}{\partial t} =
\frac{\partial}{\partial \gamma}
\left[ b( \gamma )  N( \gamma ) \right]
+ Q( \gamma )
\, .
\end{equation}

\subsection{Loss Rates} \label{sec:espect_loss}

The loss function $b ( \gamma )$ for inverse Compton scattering of the
Cosmic Microwave Background (CMB) is given by
\begin{equation} \label{eq:loss_IC}
b_{IC} ( \gamma ) =
\frac{4}{3}
\frac{ \sigma_T }{ m_e c }
\gamma^2
U_{CMB}
= 1.37 \times 10^{-20} \gamma^2 ( 1 + z )^4 \, {\rm s}^{-1}
\, ,
\end{equation}
where $\sigma_T$ is the Thomson cross-section, and $U_{CMB}$ is the
energy density in the CMB at the cluster, and $z$ is the redshift of
the cluster.
The numerical value follows from the present CMB temperature of
$T_{CMB} = 2.73$ K and the redshift evolution of the CMB energy density.
At very high energies ($\gamma \ga 10^9$), the IC losses are smaller 
because of lower Klein-Nishina cross-section (Gould 1975).

\centerline{\null}
\vskip2.55truein
\includegraphics{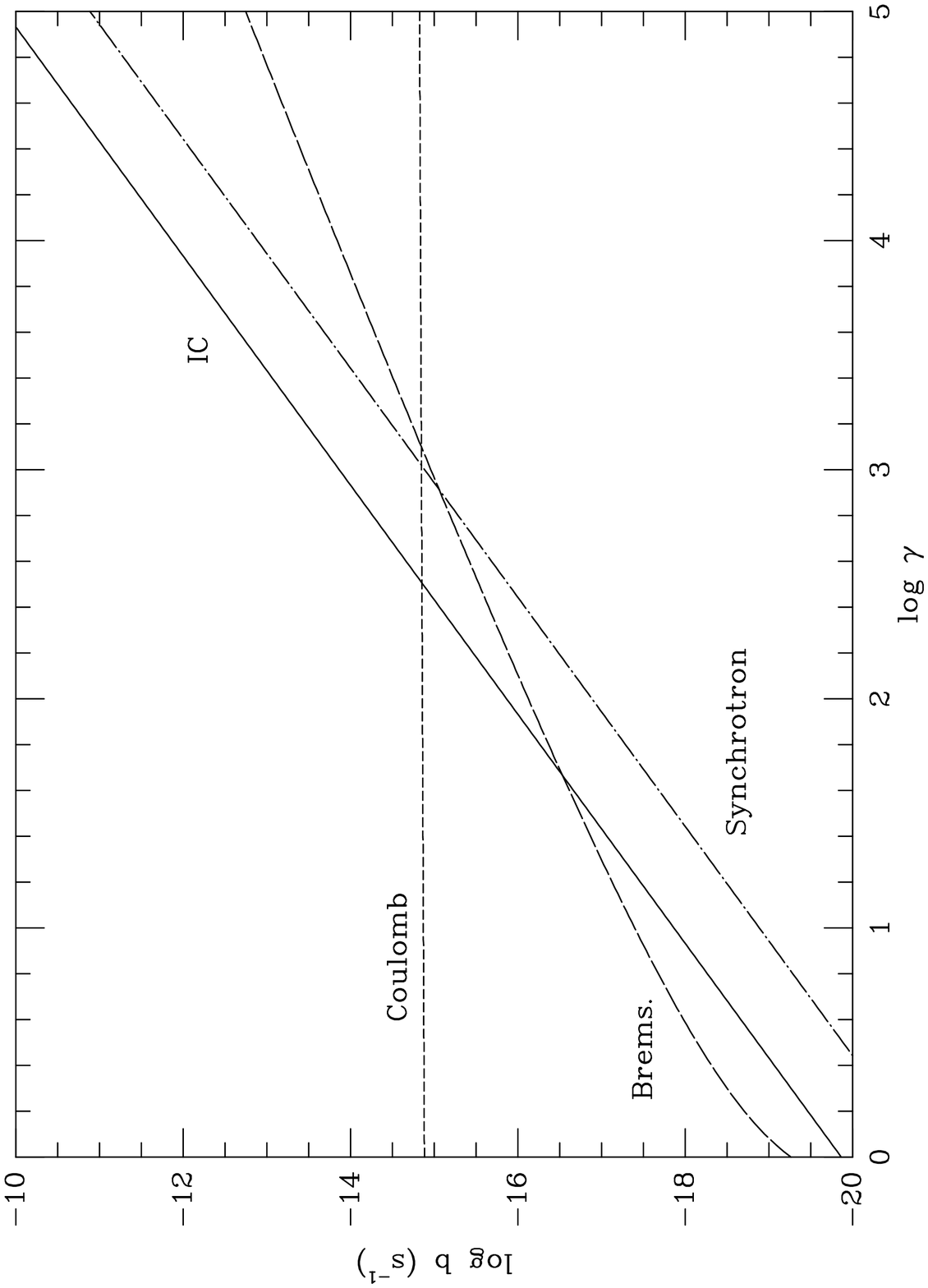}
\figcaption{
The values of the losses function $b ( \gamma )$ for inverse Compton
(IC) emission, Coulomb losses, synchrotron losses, and bremsstrahlung
losses as a function of $\gamma$.
The values assume $n_e = 10^{-3}$ cm$^{-3}$, $B = 1 \, \mu$G, and
$z = 0$.
\label{fig:losses}}

\vskip0.2truein

The expression for the loss rate due to synchrotron radiation is the
same as that in equation~(\ref{eq:loss_IC}) if one substitutes the
energy density in the magnetic field $U_B = B^2 / ( 8 \pi )$ for the
CMB energy density:
\begin{equation} \label{eq:loss_syn}
b_{syn} ( \gamma ) =
\frac{4}{3}
\frac{ \sigma_T }{ m_e c }
\gamma^2
U_{B}
= 1.30 \times 10^{-21} \gamma^2
\left( \frac{B}{1 \, \mu{\rm G}} \right)^2 \, {\rm s}^{-1}
\, ,
\end{equation}
where $B$ is the magnetic field in the ICM.
Both equations~(\ref{eq:loss_IC}) and (\ref{eq:loss_syn}) assume
$\gamma \gg 1$.
Obviously, the loss rates for IC and for synchrotron losses vary in the
same manner with $\gamma$.
The intracluster field is estimated to have a values of roughly
$B \sim 1 \, \mu$G
(e.g., Rephaeli, Gruber, \& Rothschild 1987;
Kim et al.\ 1990).
As long as the field is not much stronger than this, IC losses will
dominate over synchrotron losses, with
\begin{equation} \label{eq:IC_syn}
\frac{b_{syn}}{b_{IC}} = \frac{U_{B}}{U_{CMB}}
\approx 0.095 \,
( 1 + z )^4 \,
\left( \frac{B}{1 \, \mu{\rm G}} \right)^2
\, .
\end{equation}

The relativistic electrons will also lose energy by interactions with
the thermal plasma.
The Coulomb losses due to collisions with thermal electrons
give a loss rate which is approximately
(e.g., Rephaeli 1979)
\begin{equation} \label{eq:loss_coul}
b_{Coul} ( \gamma ) \approx
1.2 \times 10^{-12} n_e \left[ 1.0 +
\frac{\ln ( \gamma / n_e )}{75} \right] \, {\rm s}^{-1}
\, ,
\end{equation}
where $n_e$ is the thermal electron density in the cluster.
The collisions between cosmic ray electrons and thermal ions and electrons
also produce radiation through bremsstrahlung.
For the completely unscreened limit which is appropriate to a low density
plasma, the loss rate due to bremsstrahlung is given approximately by
(e.g., Blumenthal \& Gould 1970)
\begin{equation} \label{eq:loss_brem}
b_{brem} ( \gamma ) \approx
1.51 \times 10^{-16} n_e \gamma
[ \ln ( \gamma ) + 0.36] \, {\rm s}^{-1}
\, ,
\end{equation}
The numerical coefficient includes electron-ion and electron-electron
bremsstrahlung, and assumes a helium abundance by number of 10\% of
hydrogen.

The values of these loss functions as a function of $\gamma$ are given
in Figure~\ref{fig:losses} for an intracluster medium (ICM) electron density
of $n_e = 1.0 \times 10^{-3}$ cm$^{-3}$ and a typical ICM magnetic field
of $B = 1 \, \mu$G.
It is clear from this Figure and the expressions for the loss functions
that IC losses are dominant at large energies,
$\gamma \ga 200$, while Coulomb losses dominate for sufficiently small
$\gamma \la 200$ or for higher densities.
Bremsstrahlung losses are unlikely to be dominant unless the
density is higher than is typical in the bulk of the ICM (outside of 
cooling flow regions).
Similarly, synchrotron losses are unlikely to be dominant unless the
magnetic field is much stronger than 1 $\mu$G.

\subsection{Electron Lifetimes} \label{sec:espect_life}

One can define an instantaneous time scale for particle losses
by
\begin{equation} \label{eq:life_instant}
t_{loss} \equiv \frac{\gamma}{b( \gamma )} \, .
\end{equation}
Values for this loss time scale at the present time ($z = 0$) are
shown in Figure~\ref{fig:lifetime}.
The solid curve gives values assuming an average electron density of
$n_e = 10^{-3}$ cm$^{-3}$ and a magnetic field of $B = 1$ $\mu$G.
For values of the magnetic field this small or lower, synchrotron losses
are not very significant, and $t_{loss}$ is nearly independent of $B$.
The short dashed curve shows the effect of increasing the magnetic
field to $B = 5$ $\mu$G;
the losses at high energies are increased, and the loss time scales
shortened.
The dash--dot curve shows the loss time scale if the electron density is
lowered to $n_e = 10^{-4}$ cm$^{-3}$.
This reduces the losses at at low energies, and increases the loss times
there.

\centerline{\null}
\vskip2.55truein
\includegraphics{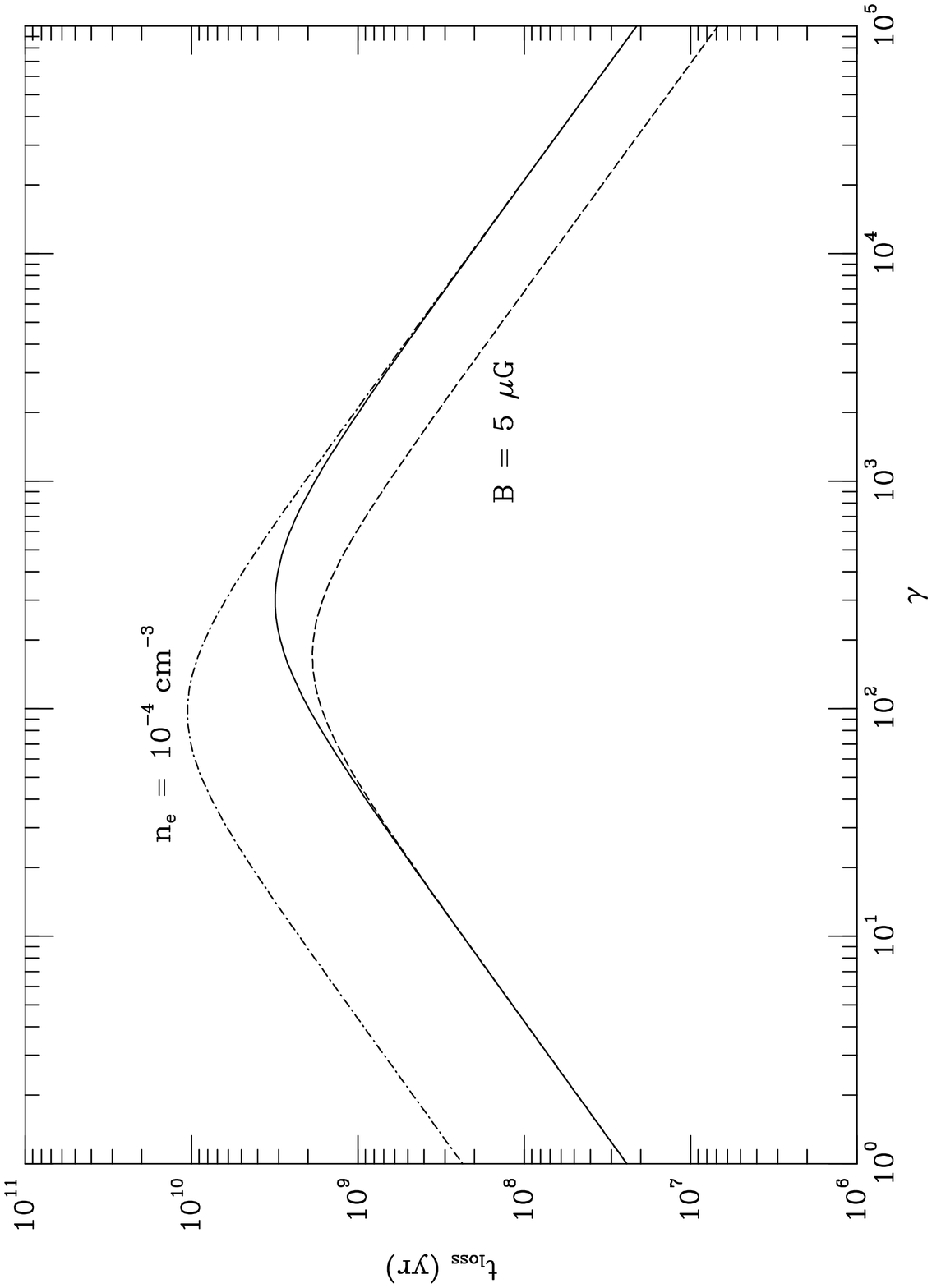}
\figcaption{
The solid curve gives the instantaneous loss time scale $t_{loss}$
(eqn.~\protect\ref{eq:life_instant}) as a function of $\gamma$
for electrons in a cluster with an electron density
of $n_e = 10^{-3}$ cm$^{-3}$ and a magnetic field of $B = 1$ $\mu$G.
The short-dash curve is for $B = 5$ $\mu$G, while the dash-dot curve
is for $n_e = 10^{-4}$ cm$^{-3}$.
\label{fig:lifetime}}

\vskip0.2truein

The characteristic features of $t_{loss}$ for parameters relevant to
clusters are that it has a maximum at $\gamma \sim 100 - 500$, and that
the maximum loss time is quite long, unless either the magnetic field
or electron density are larger than typical values for most of the
volume of a cluster.
The maximum loss times are $\sim 3 - 10$ Gyr, which is comparable to the
likely ages of clusters.
Thus, particles can accumulate with $\gamma \sim 300$ for long periods
in clusters.
In \S~\ref{sec:analytic_init} below, we will show that although electrons
at $\gamma \sim 300$ can accumulate for a significant fraction of the
Hubble time, the increase in the energy density of the CMB implies that
the accumulation period is restricted to $ z \la 1 $.

\section{Analytic Solutions} \label{sec:analytic}

A number of simple analytic solutions for the evolution of the cosmic
ray population (eq.~\ref{eq:evolution}) are possible.
I will consider two simple classes of solutions:
solutions with an initial populations of particles $N( \gamma, t_i )$
at some initial time $t_i$ or redshift $z_i$ but no subsequent injection
of additional particles [$ Q ( \gamma ) = 0$];
and solutions with no initial population 
[$N( \gamma, t_i ) = 0$] but with continual injection of particles
at a rate $ Q ( \gamma, t )$.
Since equation~(\ref{eq:evolution}) is linear in $N ( \gamma, t )$,
any solution can be written as a superposition of these two solutions.

\subsection{Initial Injection Only} \label{sec:analytic_init}

\subsubsection{General Solution} \label{sec:analytic_init_general}

First, consider a case where there is no continual source of new
particles ($Q = 0$).
Then, the population simply evolves due to the loss of energy by
individual particles.
At this point, assume that the loss function $ b ( \gamma , t )$
is an arbitrary function of $\gamma$ and time $t$.
Given an initial value for the energy of a particle $\gamma_i$ at
time $t_i$, equation~(\ref{eq:loss}) can be integrated to give the
value of $\gamma$ at a later time $t$.
Alternatively, equation~(\ref{eq:loss}) can be integrated backward
in time to give the initial energy $\gamma_i ( \gamma , t , t_i )$
at $t_i$ corresponding to a particle with energy $\gamma$ at a later
time $t$.
If there is no subsequent injection of particles, then the number
of particles is conserved if one follows their energies.
This implies that
\begin{equation} \label{eq:conserve}
\int_\gamma^\infty N ( \gamma^\prime , t ) \, d \gamma^\prime =
\int_{\gamma_i}^\infty N ( \gamma^\prime , t_i ) \, d \gamma^\prime
\, .
\end{equation}
The differential population density is then given by
\begin{equation} \label{eq:general_init}
N ( \gamma , t ) = N ( \gamma_i , t_i ) \,
\left. \frac{\partial \gamma_i}{\partial \gamma} \right\vert_t
\, .
\end{equation}

\subsubsection{Constant Loss Function} \label{sec:analytic_init_constant}

For periods of time which are short compared to the Hubble time, the
total loss function $b( \gamma )$ may be approximately independent of time.
If a particle starts (at time $t = t_i$) with a Lorentz factor
$\gamma_i$, after a time $t - t_i$ its energy will have been reduced to
$\gamma$, where
\begin{equation} \label{eq:single}
\int_{\gamma}^{\gamma_i} \frac{d \gamma^\prime}{b( \gamma^\prime )}
= ( t - t_i )
\, .
\end{equation}
Applying this energy loss to the entire electron population, one
finds that the electron spectrum evolves as
\begin{equation} \label{eq:no_source}
N ( \gamma , t ) =
N [ \gamma_i ( \gamma , t ), t_i ]
\frac{b[ \gamma_i ( \gamma , t )]}{b( \gamma )}
\, ,
\end{equation}
where 
$\gamma_i ( \gamma , t )$ is given implicitly by equation~(\ref{eq:single}).

\subsubsection{Solution at High Energies: Non-Cosmological}
\label{sec:analytic_init_highnoc}

IC losses dominate at high energies.
If we consider the evolution over time scales which are shorter
than the age of the Universe, so that the redshift $z$ is approximately
constant, then the loss function at high energies can be written as
\begin{equation} \label{eq:loss_IC_nocosmo}
b ( \gamma ) \approx b_1 \gamma^2 \qquad \gamma \gg 10^2
\, ,
\end{equation}
where the numerical constant $b_1$ is given in equation~(\ref{eq:loss_IC}).
The single particle loss equation~(\ref{eq:single}) gives 
\begin{equation} \label{eq:single_IC_nocosmo}
\gamma_i = \frac{\gamma}{1 - \frac{\gamma}{\gamma_{max}}}
\, .
\end{equation}
After the passage of a time $\Delta t \equiv ( t - t_i )$,
all of the electrons with energies above $\gamma_{max}$ will have been
removed, where
\begin{equation} \label{eq:gmax_IC_nocosmo}
\gamma_{max} = \frac{1}{b_1 \Delta t}
\, .
\end{equation}
Thus, the resulting electron distribution is related to the initial
distribution by
\begin{equation} \label{eq:distr_IC_nocosmo}
N ( \gamma, t ) = \left\{
\begin{array}{cl}
\frac{N [ \gamma / ( 1 - \gamma / \gamma_{max} ), t_i ]}
{(1 - \gamma / \gamma_{max} )^2} &
\gamma < \gamma_{max} \\
& \\
0 & \gamma \ge \gamma_{max} \\
\end{array}
\right.
\end{equation}
If the initial particle distribution was a power-law with
$N ( \gamma_i , t_i ) = N_1 \gamma_i^{-p_0}$, then the resulting final
distribution is
\begin{equation} \label{eq:distp_IC_nocosmo}
N ( \gamma, t ) = \left\{
\begin{array}{cl}
N_1 \gamma^{-p_0} ( 1 - \gamma / \gamma_{max} )^{p_0-2} & \gamma < \gamma_{max} \\
& \\
0 & \gamma \ge \gamma_{max} \\
\end{array}
\right.
\end{equation}

These equations also apply if there if a significant magnetic field in the
cluster, since the energy losses from synchrotron emission have the same
dependence on $\gamma$ as those for IC scattering.
The value of $b_1$ would be increased to include synchrotron losses
according to equations~(\ref{eq:loss_IC}) and (\ref{eq:loss_syn}).

\subsubsection{Solution at High Energies: Cosmological}
\label{sec:analytic_init_highc}

These expressions do not apply to the evolution of the electron energy
spectrum over cosmological time scales because of the variation in
the energy density of the CMB with redshift $z$.
As a result, the loss rate due to IC emission varies with electron
energy and redshift as
(eq.~\ref{eq:loss_IC})
\begin{equation} \label{eq:loss_IC_1}
b_{IC} ( \gamma , z ) = b_1 \gamma^2 ( 1 + z )^4 \, .
\end{equation}
The equation for the evolution of the energy of a single particle subject
to IC losses is then
\begin{equation} \label{eq:dgamma}
\frac{ d \gamma}{\gamma^2} = - b_1 ( 1 + z )^4 \, dt
\, .
\end{equation}
The age of the Universe $t$ and redshift $z$ are related implicitly
by the equation of cosmic dynamics
(e.g., Weinberg 1972),
\begin{equation} \label{eq:cosmic_dyn}
t = \frac{1}{H_0} \, \int_0^{(1 + z)^{-1}}
\left( 1 - 2 q_0 + \frac{2 q_0}{x} \right)^{-1/2} \, d x
\, ,
\end{equation}
here $H_0$ is the Hubble constant and $q_0$ is the deceleration parameter
at the present time.

After some algebra, equation~(\ref{eq:dgamma}) can be integrated to yield
the current $\gamma$ for an electron which had an initial energy of $\gamma_i$
at a redshift $z_i$:
\begin{equation} \label{eq:evol_cosmic}
\frac{1}{\gamma} - \frac{1}{\gamma_i} =
\frac{b_1}{H_0} \, f( q_0 , z_i ) \, ,
\end{equation}
where the function $f( q_0 , z )$ is defined by
\begin{eqnarray}
f( q_0 , z ) & \equiv & \frac{1}{15 q_0^3}
\left[ \sqrt{1 + 2 q_0 z}
\left( 2 - 10 q_0 + 15 q_0^2 - 2 q_0 z \right. \right. \nonumber \\
& & \qquad \mbox{} \left. \left. + 10 q_0^2 z + 3 q_0^2 z^2 \right)
-2 + 10 q_0 -15 q_0^2 \right] \, . \label{eq:fcosmic}
\end{eqnarray}
For $q_0 = 0$, the limit of equation~(\ref{eq:fcosmic}) is
\begin{equation} \label{eq:fcosmic0}
f( 0 , z ) = z \left( 1 + z + \frac{z^2}{3} \right) \, .
\end{equation}

\centerline{\null}
\vskip2.55truein
\includegraphics{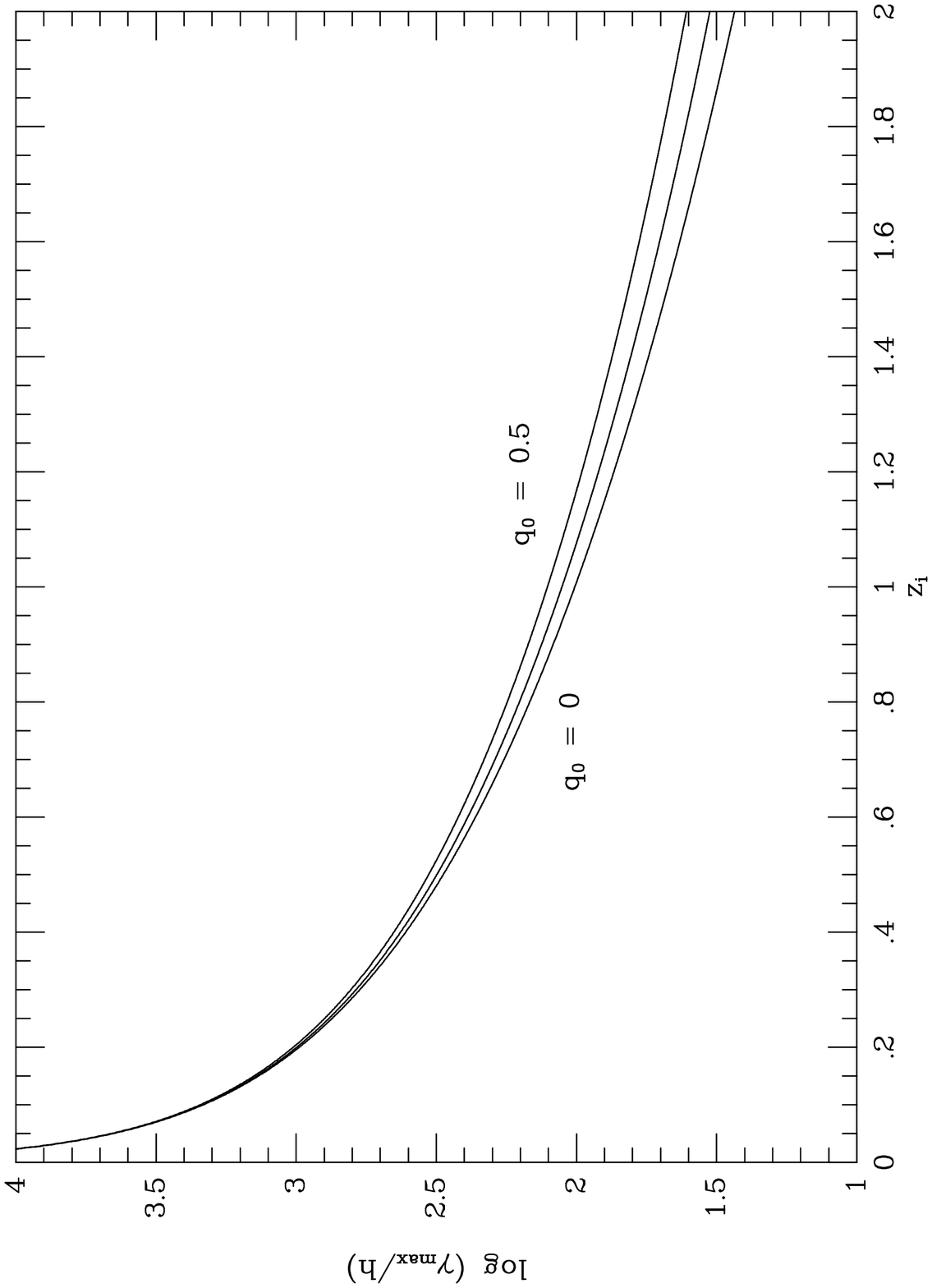}
\figcaption{
The present values of the maximum remaining electron Lorentz factor
$\gamma_{max}$ is shown as a function of the redshift $z_i$ at which 
the electrons were last injected in the cluster.
Only IC losses are included.
Curves are shown for three cosmological models with values of the cosmic
deceleration parameter of $q_0 = 0$, 0.2, and 0.5.
The values of $\gamma_{max}$ are scaled to the value of the present Hubble
constant, with $h \equiv ( H_0 /100$ km s$^{-1}$ Mpc$^{-1}$).
\label{fig:gmax}}

\vskip0.2truein

The cosmological expression for the maximum remaining energy of electrons
subject to IC losses is then
\begin{equation} \label{eq:gmax_IC_cosmo}
\gamma_{max} = \frac{H_0}{b_1} \, \frac{1}{f( q_0 , z_i )}
\, .
\end{equation}
Figure~\ref{fig:gmax} shows values of the maximum Lorentz factor
as a function of the redshift $z_i$ for cosmological models with
values of $q_0 = 0$, 0.2, and 0.5.
The values of $\gamma_{max}$ are scaled by $h \equiv ( H_0 /100$ km s$^{-1}$
Mpc$^{-1}$).

\subsubsection{Solution at High Energies with Synchrotron Losses}
\label{sec:analytic_init_highb}

If the ICM magnetic field is much stronger than is currently believed,
synchrotron losses might be important for high energy electrons.
Then, the expressions for the cosmological evolution of an initial electron
population at high energies would be modified in a way which would depend
on the temporal variation of the magnetic field.
If the magnetic field were constant, then the losses at high energies could
be written as
\begin{equation} \label{eq:loss_highE}
b ( \gamma , z ) \approx \gamma^2
\left[ b^{IC}_1 ( 1 + z )^4 + b^{syn}_1 \right] \, ,
\end{equation}
where the values of $b^{IC}_1$ and $b^{syn}_1$ can be found from
equations~(\ref{eq:loss_IC}) and (\ref{eq:loss_syn}).
Then the evolution of the energy of a single particle at high energies is
given by
\begin{equation} \label{eq:evol_cosmic_B}
\frac{1}{\gamma} - \frac{1}{\gamma_i} =
\frac{1}{H_0}
\left\{ b^{IC}_1 f( q_0 , z_i ) + b^{syn}_1 [ g(q_0, 0) - g( q_0, z_i )]
\right\} \, ,
\end{equation}
where $g( q_0 , z )$ is the integral on the right side of
equation~(\ref{eq:cosmic_dyn}).
This function is most easily given in the standard parametric form
(e.g., Weinberg 1972)
\begin{equation} \label{eq:cosmology}
g ( q_0 , z ) = \left\{
\begin{array}{lll}
( 1+z)^{-1} & & q_0 = 0 \\
&& \\
\frac{q_0 [ \sinh ( \theta ) - \theta ]}{( 1 - 2 q_0 )^{3/2}} &
\cosh ( \theta ) = 1 + \frac{1 - 2 q_0}{q_0} \, \frac{1}{1+z} &
0 < q_0 < 1/2 \\
&& \\
\frac{2}{3} \, ( 1+z)^{-3/2} & & q_0 = 1/2 \\
&& \\
\frac{q_0 [ \theta - \sin ( \theta )]}{(2 q_0  - 1)^{3/2}} &
\cos ( \theta ) = 1 - \frac{2 q_0 - 1}{q_0} \, \frac{1}{1+z} &
1/2 < q_0 \\
\end{array}
\right.
\end{equation}
Then, the upper cutoff in the electron energy distribution is given by
\begin{equation} \label{eq:gmax_B_cosmo}
\gamma_{max} = \frac{H_0}{b^{IC}_1 f( q_0 , z_i ) +
b^{syn}_1 [g(q_0, 0 ) - g( q_0 , z_i )] }
\, .
\end{equation}

The resulting particle distributions are still given by
equations~(\ref{eq:distr_IC_nocosmo}) or
(\ref{eq:distp_IC_nocosmo}), with the correct value for $\gamma_{max}$.
If one wants to calculate the particle distribution at an intermediate
redshift z ($ 0 < z < z_i$) rather than at present ($z = 0$), then
one substitutes $[ f ( q_0 , z_i ) - f( q_0 , z ) ]$ for $f ( q_0 , z_i )$
and $g ( q_0 , z )$ for $g ( q_0 , 0 )$ in
equations~(\ref{eq:evol_cosmic}),
(\ref{eq:gmax_IC_cosmo}),
(\ref{eq:evol_cosmic_B}), and
(\ref{eq:gmax_B_cosmo}).

\subsubsection{Solution at Low Energies}
\label{sec:analytic_init_low}

At low electron energies, the dominant losses will be due to Coulomb
collisions.
Coulomb losses have the property that they vary only rather slowly
(logarithmically) with $\gamma$
(eq.~\ref{eq:loss_coul}).
Here, this slow variation is ignored, and the Coulomb losses are
written as $b ( \gamma , t ) = b_{Coul} \approx b_C$, where $b_C$ is
a constant given by the value in equation~(\ref{eq:loss_coul}) for
some appropriate value of $\gamma$, say $\gamma \sim 10^2$.
Then, equation~(\ref{eq:single}) for the change in $\gamma$ for a single
particle gives
\begin{equation} \label{eq:evol_low}
\gamma = \gamma_i - \gamma_{low} \, ,
\end{equation}
where the characteristic value $\gamma_{low}$ is defined by
\begin{equation} \label{eq:glow}
\gamma_{low} \equiv b_{C} \Delta t \, .
\end{equation}
The equation for the evolution of the electron population
(eq.~\ref{eq:no_source}) reduces to
\begin{equation} \label{eq:low_energy_soln1}
N ( \gamma , t ) \approx N [ \gamma + \gamma_{low} , t_i ]
\, .
\end{equation}
For a short time $\Delta t \ll \gamma / b_{C}$,
the population is nearly the initial one.
However, the Coulomb losses at low electron energies will tend to be rapid.
Thus, after a time $\Delta t \gg \gamma / b_{C}$,
the population will approach
\begin{equation} \label{eq:low_energy_soln2}
N ( \gamma , t ) \approx N [ \gamma_{low} , t_i ]
\, .
\end{equation}
That is, the population at low energies will vary with time, but will
be nearly independent of $\gamma$,
\begin{equation} \label{eq:low_energy_soln}
N( \gamma ) \approx N_{low} (t)  = {\rm constant }
\, .
\end{equation}

\subsection{Steady Injection Only} \label{sec:analytic_steady}

As an alternative to the case where there is an initial population of particles
but no continual source, we consider the case where there is
continuous injection of particles at a constant rate $Q ( \gamma )$,
but where there is no initial electron population.

\subsubsection{Steady-State Solutions} \label{sec:analytic_steady_steady}

At energies where the time scale for losses by particles $t_{loss}$ is much
shorter than the age of the cluster, the population will tend towards
a steady-state distribution.
From Figure~\ref{fig:lifetime}, this is likely to occur at high energies
($\gamma \ga \gamma_{max}$) due to rapid IC and synchrotron losses.
It may also occur at low energies ($\gamma \la \gamma_{low}$)
due to rapid Coulomb losses.
If $t_{loss} \equiv [ \gamma / b ( \gamma ) ] \ll \Delta t$, then
it is likely that
$\partial N ( \gamma , t ) / \partial t \ll
\partial [ b( \gamma , t )  N( \gamma , t ) ] / \partial \gamma$.
In steady-state ($\partial / \partial t = 0$), the equation for the
evolution of the electron population (eq.~\ref{eq:evolution}) becomes
\begin{equation} \label{eq:steady_evol}
\frac{d}{d \gamma}
\left[ b( \gamma )  N( \gamma ) \right] = - Q( \gamma )
\, .
\end{equation}
If there is no flux of particles from infinite energies and if
$\lim_{\gamma \rightarrow \infty} b( \gamma ) N( \gamma ) \rightarrow 0$,
then the steady-state solution is given by
\begin{equation} \label{eq:steady}
N ( \gamma ) = \frac{1}{b( \gamma )}
\int_\gamma^\infty Q ( \gamma^\prime ) \, d \gamma^\prime
\, .
\end{equation}

It is useful to consider a source which is a power-law function of the
electron energy, with
\begin{equation} \label{eq:source}
Q( \gamma ) = Q_1 \gamma^{-p}
\, ,
\end{equation}
with $p > 1$.
Then, at high energies where IC and synchrotron losses dominate and
we can write the loss function as $b ( \gamma ) \approx b_1 \gamma^2$, the
resulting electron distribution is
\begin{equation} \label{eq:steady_high}
N( \gamma ) = \frac{Q_1}{b_1 (p - 1)} \, \gamma^{-(p+1)} \, ;
\end{equation}
that is, the resulting electron distribution is one power steeper than
the source.
At low energies where Coulomb losses dominate, the losses depend only
logarithmically on the electron energy.
If we ignore this slow dependence and take $b ( \gamma ) \approx b_C$ (a
constant), then the resulting steady-state distribution is
\begin{equation} \label{eq:steady_low}
N( \gamma ) = \frac{Q_1}{b_C (p - 1)} \, \gamma^{-(p-1)} \, ;
\end{equation}
that is, the resulting electron distribution is one power flatter than
the source.
These classical results were derived first by Ginzburg
(e.g., Ginzburg \& Syrovatskii 1964).

\subsubsection{Self-Similar Solutions} \label{sec:analytic_steady_ss}

For simple approximations for the loss function and simple expressions
for the rate of injection $Q ( \gamma )$, one can often find self-similar
solutions which connect the steady-state population with non-steady
populations which are time dependent.
In these solutions, the overall normalization of the electron population
varies with time, but the shape of the energy spectrum remains the same if
$\gamma$ is scaled by some characteristic value.

At high energies where IC and synchrotron losses dominate,
one can approximate the loss function by
equation~(\ref{eq:loss_IC_nocosmo}), at least at low redshifts.
The characteristic value of $\gamma$ at high energies is $\gamma_{max}$
(eq.~\ref{eq:gmax_IC_nocosmo}).
Then, if the particle injection follows a power-law distribution
(eq.~\ref{eq:source}), the self-similar solution for the electron
population is given by
\begin{equation} \label{eq:self_similar_IC}
N( \gamma, t ) = \frac{Q_1}{b_1 (p - 1)} \, \gamma^{-(p+1)}
\left\{
\begin{array}{cl}
1 - ( 1 - \gamma / \gamma_{max} )^{p-1} &
\gamma < \gamma_{max} \\
& \\
1 & \gamma \ge \gamma_{max} \\
\end{array}
\right. \, .
\end{equation}

At low energies where Coulomb losses dominate, the loss rate in
nearly independent of $\gamma$, $b ( \gamma ) \approx b_C$.
Then, the characteristic value of the $\gamma$ is $\gamma_{low}$
(eq.~\ref{eq:glow}).
For a power-law spectrum of injected particles (eq.~\ref{eq:source}), the
self-similar solution for the electron population is given by
\begin{equation} \label{eq:self_similar_low}
N( \gamma, t ) = \frac{Q_1}{b_C (p - 1)} \, \gamma^{-(p-1)}
\left[
1 - \left( 1 + \gamma_{low} / \gamma \right)^{-(p-1)}
\right] \, .
\end{equation}

At energies where losses are not important ($\gamma \ll \gamma_{max}$ in
equation~[\ref{eq:self_similar_IC}] and $\gamma \gg \gamma_{low}$
in equation~[\ref{eq:self_similar_low}]), the two self-similar
solutions are identical, $ N ( \gamma , t ) \approx Q ( \gamma ) t$.
That is, the population is just the sum of the injected particles at
that energy, as expected if losses have not affected the energies of
the particles.
If $\gamma_{low} \ll \gamma_{max}$, then the two solutions can simply
be joined in this central region.
With some algebra, one finds that this solution can be written as
\begin{eqnarray}
N( \gamma, t ) \approx \frac{Q_1 \gamma^{-p} t}{p - 1} \,
\left( \frac{\gamma_{low}}{\gamma} + \frac{\gamma}{\gamma_{max}} \right)^{-1}
\hspace{1.5truein}
&& \nonumber \\
\times \left\{ 2 - 
\left[ \frac12
\left( 1 - \frac{\gamma}{\gamma_{max}} +
\left\vert 1 - \frac{\gamma}{\gamma_{max}} \right\vert \right)
\right]^{p-1}
- \left( 1 + \frac{\gamma_{low}}{\gamma} \right)^{-(p-1)} \right\}
\, . && \label{eq:self_similar}
\end{eqnarray}
This solution applies for all $\gamma$ as long as
$\gamma_{low} \ll \gamma_{max}$.

\section{Source Function for Electrons} \label{sec:source}

\subsection{Injected Electron Spectrum} \label{sec:source_spectrum}

Models will be calculated which include an initial
populations of relativistic electrons and/or continuous injection
of relativistic electrons during the history of the cluster.
Although the models don't depend on the details of the processes
which accelerate relativistic electrons in clusters, the basic
assumption is that these particles result from shock or turbulent
acceleration.
The shocks and turbulence might be associated with the hydrodynamical
evolution of the ICM;
for example, the initial population might be due to an accretion
shock during the formation of the cluster, and semi-continuous particle
injection might result from subcluster mergers.
Alternatively, the shocks might be due to the expansion of radio sources
within the ICM.

The condition for the shocks due to the hydrodynamics of the intracluster
medium are similar to condition in Galactic supernova remnants (SNRs),
except that the length scales are much larger and the densities are smaller.
Over a significant range of relativistic energies, there is considerable
theoretical and observational evidence that such shock acceleration
produces particles with a power-law distribution
(e.g., Blandford \& Eichler 1987;
Jones \& Ellison 1991).
Thus, we will consider models in which the initial electron spectrum
at $t = t_i$ is a single power-law,
\begin{equation} \label{eq:initial_pw1}
N( \gamma , t_i ) = N_1 \gamma^{-p_0}
\, ,
\end{equation}
and/or the continuous source of injected electrons is a power-law,
\begin{equation} \label{eq:injection_pw1}
Q( \gamma , t ) = Q_1 ( t ) \gamma^{-p}
\, .
\end{equation}
For typical strong shock compressions, one expects values of
$p = p_0 \approx 2.3$
(e.g., Blandford \& Eichler 1987;
Jones \& Ellison 1991).

One complication in applying the standard results on shock acceleration
to the ICM is that we need to determine the population of relatively low
energy cosmic ray electrons ($\gamma \sim 300$)
in order to calculate their EUV emission.
Such particles have kinetic energies ($\sim$150 MeV) for which electrons
are relativistic, but protons and other ions are not.
There are a number of simple arguments which suggest that the energy
spectrum of superthermal but subrelativistic particles is a flatter
power-law than for fully relativistic particles
(Bell 1978b).
The suggestion is that the {\it momentum} spectrum of particles accelerated
in shocks is the same power-law for both
relativistic and nonrelativistic particles and for both ions and
electrons.
That is, $Q ( P ) d P \propto P^{-p} dP$, where $P$ is the particle momentum
and $Q( P ) d P$ gives number of particles accelerated with momenta between
$P$ and $P + d P$ per unit time.
For highly relativistic particles $E = P c$, but for non-relativistic particles the kinetic energy $E = P^2/ 2 m$, where $m$ is their mass.
Under the same assumptions which result in a shock acceleration energy
spectrum of highly relativistic particles with $Q(E) \propto E^{-p}$,
the energy spectrum for superthermal but subrelativistic particles is
$Q(E) \propto E^{-(p+1)/2}$.
For a typical relativistic injection spectrum with $p \approx 2.3$,
the subrelativistic spectrum would have a lower exponent
$p_l \approx 1.65$.
At energies $E \sim 100$ MeV, this would only affect the shock acceleration
spectrum of ions.
Bell (1978b) suggests that would result in the total number of accelerated
electrons and ions (at energies above $m_e c^2$) being nearly equal,
while the number of ions with $ E \gg m_p c^2$ would be larger than
that of electrons at the same energies.

However, it is also possible the the spectrum of shock accelerated
electrons would also be affected.
Thus, we will also consider models in which the initial electron spectrum
at $t = t_i $ is a broken power-law,
\begin{equation} \label{eq:initial_pw2}
N( \gamma , t_i ) = N_1 \times \left\{
\begin{array}{cl}
\gamma^{-p_l} & \gamma < \gamma_{br} \\
& \\
\gamma_{br}^{-p_l} \left( \frac{\gamma}{\gamma_{br}} \right)^{-p_0} &
\gamma \ge \gamma_{br} \\
\end{array}
\right.
\end{equation}
and/or the continuous source of injected electrons is a broken power-law,
with
\begin{equation} \label{eq:injection_pw2}
Q( \gamma , t ) = Q_1 ( t ) \left\{
\begin{array}{cl}
\gamma^{-p_l} & \gamma < \gamma_{br} \\
& \\
\gamma_{br}^{-p_l} \left( \frac{\gamma}{\gamma_{br}} \right)^{-p} &
\gamma \ge \gamma_{br} \\
\end{array}
\right.
\end{equation}

\subsection{Normalization of the Electron Population}
\label{sec:source_normalization}

Equation~\ref{eq:evolution} for the evolution of the population is linear
in $N ( \gamma )$, so the solution can be rescaled by increasing or decreasing
$N ( \gamma , t )$, the initial value $N ( \gamma , t_i )$, and the
electron injection rate $ Q ( \gamma , t)$ by any constant factor.
Thus, the normalization of the solution is arbitrary.
I have chosen an arbitrary normalization;
however, it is at least somewhat consistent with energetic and stability
arguments concerning the ICM medium, with the observed EUV fluxes of
clusters, and with the observed properties of shock acceleration in
SNRs in our Galaxy.
If the observed EUV radiation from clusters results from IC scattering of
CMB photons by cosmic ray electrons, the observed EUV luminosities imply
that clusters contain cosmic ray electrons with $\gamma \ga 300$
which have a total energy of the order of $E_{CR,e} \sim 10^{62}$ ergs.
It is also unlikely that the cosmic ray energy significantly exceeds the
thermal energy of the ICM, since this would likely result in the
ICM being expelled from the clusters.
Rich clusters contain approximately $10^{14} \, M_\odot$ of hot gas at
a typical temperature of $7 \times 10^7$ K, which implies that the
thermal energy content of the gas is approximately
$E_{gas} \sim 4 \times 10^{63}$ ergs.
Thus, if the cosmic ray electrons had a total kinetic energy
of $E_{CR,e} \sim 10^{62}$ ergs, this would represent approximately 3\%
of the thermal energy content of the ICM.
Even when one includes the contribution from cosmic ray ions, this
ratio is likely to remain of order unity or smaller.
Finally, we note that the radio observations of Galactic SNR imply
that shock acceleration produces cosmic ray electrons which contain
at least a few per cent of the shock energy
(e.g., Blandford \& Eichler 1987).
Since essentially all of the thermal energy content of the ICM is the
result of shock heating (\S~\ref{sec:intro} above), these shocks should
also have accelerated relativistic electrons containing at least a
few per cent of the thermal energy content of the intracluster medium.

I have chosen to normalize the electron spectra so that the
total amount of kinetic energy injected in electrons with $\gamma \ge 300$
is $E^{tot}_{CR,e} = 10^{63}$ ergs.
In the models which seem most likely to represent real, present day
clusters, this leads to present day electron populations with
$E_{CR,e} \sim 10^{62}$ ergs.

\subsection{Self-Similar Cluster Accretion Shock}
\label{sec:source_bertschinger}

In most of the models with continual injection of particles, we assume
for simplicity that the injection rate is constant over the time that
it acts, $ \partial Q ( \gamma , t ) / \partial t = 0 $.
As an alternative, we consider a simple model for the cosmological
evolution of $ Q ( \gamma , t )$.
Bertschinger (1985) has given a simple, self-similar solution for the
secondary accretion of intracluster gas (with or without associated
dark matter) onto a previously collapse cluster core.
This solution applies only in an Einstein-de Sitter Universe,
with $\Omega = 1$ and $\Lambda = 0$.
In the Bertschinger solution,
the accretion shock radius expands with time as $r_s \propto t^{8/9}$, and
the preshock density varies with the average cosmological density
$\rho \propto t^{-2}$.
The shock velocity obviously varies at
$v_s \propto r_s / t \propto t^{-1/9}$.
In this solution, the preshock gas is cold and the accretion shock is
always very strong.
Thus, rate of shock energy input varies as
$\dot{E}_{shock} = (1/2) \rho v_s^3 4 \pi r_s^2 \propto t^{-5/9}$.

If we assume that a given fraction of the shock energy goes into
accelerating the relativistic electrons, then it is useful to give the
rate of shock energy input at any time in terms of the total shock energy
over the life time of the cluster.
Integration of the shock energy equation leads to the following expression
for the accretion shock energy input in the Bertschinger self-similar
solution:
\begin{equation} \label{eq:injection_bert}
\frac{\dot{E}_{shock} (z) }{E_{shock}} =
\frac{4}{9} \, \frac{1}{t_o}
\frac{( 1 + z )^{5/6}}{1 - ( 1 + z_i )^{-2/3}} 
\, ,
\end{equation}
where $t_o$ is the present age of the Universe, $z$ is the redshift at
the time when the shock energy is being evaluated, and $z_i$ is the
redshift at which the cluster formed.

%
% table of numerical models 
%
% SUBMISSION
% \begin{table}[thb]
% \small
% \caption{\hfil Numerical Models for the Electron Spectrum \label{tab:models} \hfil}
% PREPRINT
\begin{table*}[htb]
\small
\tabcaption{\hfil Numerical Models for the Electron Spectrum \label{tab:models} \hfil}
\begin{center}
\begin{tabular}{lccccccccccccl}
\tableline
\tableline
&$z_i$&$z_s$&$H_0$&$n_e$&$B$&Initial&$p_0$&Cont.&$p$&$\gamma_{br}$&$p_l$&$F_{inj}$&\cr
Model&&&(km s$^{-1}$ Mpc$^{-1}$)&(cm$^{-3}$)&($\mu$G)&Pop.&&Inj.&&&&(\%)&Comment\cr
\tableline
 1&2.00&2.00&\phn65&0.001\phn&0.3&N&   &Y&2.3&    &   &100&Steady injection\cr
 2&1.00&1.00&\phn65&0.001\phn&0.3&N&   &Y&2.3&    &   &100&Steady injection\cr
 3&0.50&0.50&\phn65&0.001\phn&0.3&N&   &Y&2.3&    &   &100&Steady injection\cr
 4&0.30&0.30&\phn65&0.001\phn&0.3&N&   &Y&2.3&    &   &100&Steady injection\cr
 5&0.10&0.10&\phn65&0.001\phn&0.3&N&   &Y&2.3&    &   &100&Steady injection\cr
 6&0.01&0.01&\phn65&0.001\phn&0.3&N&   &Y&2.3&    &   &100&Steady injection\cr
 7&2.00&2.00&\phn65&0.0001   &0.3&N&   &Y&2.3&    &   &100&Model 1, but lower density\cr
 8&2.00&2.00&\phn65&0.000\phn&5.0&N&   &Y&2.3&    &   &100&Model 1, but stronger magnetic field\cr
 9&0.01&    &\phn65&0.001\phn&0.3&Y&2.3&N&   &    &   &0&Initial pop.\ only\cr
10&0.10&    &\phn65&0.001\phn&0.3&Y&2.3&N&   &    &   &0&Initial pop.\ only\cr
11&0.30&    &\phn65&0.001\phn&0.3&Y&2.3&N&   &    &   &0&Initial pop.\ only\cr
12&0.50&    &\phn65&0.001\phn&0.3&Y&2.3&N&   &    &   &0&Initial pop.\ only\cr
13&1.00&    &\phn65&0.001\phn&0.3&Y&2.3&N&   &    &   &0&Initial pop.\ only\cr
14&0.30&    &\phn65&0.0001   &0.3&Y&2.3&N&   &    &   &0&Model 11, but lower density\cr
15&0.30&    &   100&0.001\phn&0.3&Y&2.3&N&   &    &   &0&Model 11, but larger $H_0$\cr
16&0.30&    &\phn65&0.001\phn&0.0&Y&2.3&N&   &    &   &0&Model 11, but no magnetic field\cr
17&0.30&    &\phn65&0.001\phn&1.0&Y&2.3&N&   &    &   &0&Model 11, but stronger magnetic field\cr
18&0.30&    &\phn65&0.001\phn&3.0&Y&2.3&N&   &    &   &0&Model 11, but stronger magnetic field\cr
19&0.30&    &\phn65&0.001\phn&5.0&Y&2.3&N&   &    &   &0&Model 11, but stronger magnetic field\cr
20&2.00&2.00&\phn65&0.001\phn&0.3&N&   &Y&2.3&    &   &100&Bertschinger model\cr
21&2.00&2.00&\phn65&0.001\phn&0.3&N&   &Y&2.3&2000&1.3&100&Broken power-law injection\cr
22&0.30&    &\phn65&0.001\phn&0.3&Y&2.3&N&   &2000&1.3&0&Broken power-law initial pop.\cr
23&0.05&0.05&\phn65&0.001\phn&0.3&N&   &Y&2.3&    &   &100&Steady injection\cr
24&0.30&0.05&\phn65&0.001\phn&0.3&Y&2.3&Y&2.3&    &   & 50&Both init.\ pop.\ \& current inj.\cr
25&0.30&0.05&\phn65&0.001\phn&0.3&Y&2.3&Y&2.3&    &   & 25&Both init.\ pop.\ \& current inj.\cr
26&0.30&0.05&\phn65&0.001\phn&0.3&Y&2.3&Y&2.3&    &   & 10&Both init.\ pop.\ \& current inj.\cr
27&0.30&0.05&\phn65&0.001\phn&0.3&Y&2.3&Y&2.3&    &   &  1&Both init.\ pop.\ \& current inj.\cr
28&0.30&0.05&\phn65&0.001\phn&0.3&Y&2.3&Y&2.3&    &   &0.1&Both init.\ pop.\ \& current inj.\cr
\tableline
\end{tabular}
\end{center}
% SUBMISSION
% \end{table}
% PREPRINT
\end{table*}

\section{Numerical Models} \label{sec:numerical}

\subsection{Techniques} \label{sec:numerical_techniques}

A number of numerical models for the integrated energy spectra of cosmic
ray electrons in clusters of galaxies have been calculated.
The models were constructed so that the present day electron populations
could be determined up to at least $\gamma = 10^5$.
In models without continual electron injection, the present day electron
populations are generally cut-off at $\gamma_{max} < 10^5$.
In the models with continual electron injection, the electron population
is in steady-state above $\gamma = 10^5$ in all of the models
considered, and the populations can easily be determined from
equations~(\ref{eq:steady}) or (\ref{eq:steady_high}).
Also, the rapid loss rates for high energy electrons with $\gamma > 10^5$
mean that numerical solutions for these electrons are time consuming.

For models with an initial electron population at redshift $z_i$ but
no subsequent particle injection, the population was calculated by
following the energy losses of individual particles.
That is, equation~(\ref{eq:loss}) was integrated backward in time to
determine the initial Lorentz factor $\gamma_i$ corresponding to the
present day $\gamma$ for a large number of particles with energies
in the range $1 < \gamma \le 10^5$.
The loss equation was integrated using the
Bulirsch-Stoer technique with adaptive step-size control
(e.g., Press et al.\ 1986).
The equation of cosmic dynamics (eq.~\ref{eq:cosmic_dyn}) was used to
relate redshift and co-moving time.
All of the loss processes discussed in \S~\ref{sec:espect_loss} were
included.
Then, the present day electron spectrum was determined from particle
conservation through equation~(\ref{eq:general_init}).

For models with continual injection but no initial population, the
equation for the evolution of the electron energy spectrum
(eq.~\ref{eq:evolution}) was integrated with a finite-difference technique
on a fixed grid of energies covering the range $1 < \gamma \le 10^5$.
I found that the energy spectrum at very high energies was well-represented
by the self-similar solution (eq.~\ref{eq:self_similar_IC}).
I assumed this solution applied at high energies beyond the upper end
of the grid, and it was used to determine the flux of particles into
the grid from higher energies.
The time step was regulated so that the flux of particles into each grid
cell from higher energies was a small fraction ($<$1\%) of the population
of that cell, and so that the redshift step was also very small.

Because the equation for the evolution of the electron population
is linear (eq.~\ref{eq:evolution}), models with both an initial
populations and subsequent particle injection were created by the
superposition of solutions with only and initial population and
only subsequent injection.

Unless otherwise noted, the average thermal electron
density in the ICM was taken to be $n_e = 0.001$ cm$^{-3}$,
the Hubble constant was $H_0 = 65$ km s$^{-1}$ Mpc$^{-1}$, and the
magnetic field was $B = 0.3$ $\mu$G.
In all of the models, the deceleration parameter was $q_0 = 0.5$.

In all of the models, there is a population of particles with
$\gamma \approx 1$, which represent initially relativistic electrons which
have returned to the thermal population through energy losses.
Since the number of cosmic ray electrons is small compared to the
number of thermal electrons, this population is probably not important.

% \placetable{tab:models}

Some of the assumed properties of the numerical models are summarized
in Table~\ref{tab:models}.
The models start at a redshift of $z_i$.
In models with continual particle injection, this starts at $z_s$.
The value of the Hubble constant in the models is given in column 4.
The average thermal electron density $n_e$ and magnetic field $B$ are given
in columns 5 and 6.
A ``Y'' in the columns labeled ``Initial'' or ``Cont.'' indicates whether
the model includes an initial population of relativistic electron
at $z_i$ and/or whether there is continual particle injection since $z_s$.
The values of $p_0$, $p$, $\gamma_{br}$, and $p_l$ indicate the form
of the power-law or broken power-law injection spectrum, according to
equations~(\ref{eq:initial_pw1})--(\ref{eq:injection_pw2}).
If no value is given for $\gamma_{br}$, the injection spectrum is a
single power-law (eqs.~\ref{eq:initial_pw1}--\ref{eq:injection_pw1}).
For models with both an initial population and continual particle
injection, the value of $F_{inj}$ is the fraction of the total particle
energy which is contributed by the continual particle injection.
The last column gives a comment on the model.

\centerline{\null}
\vskip2.55truein
\includegraphics{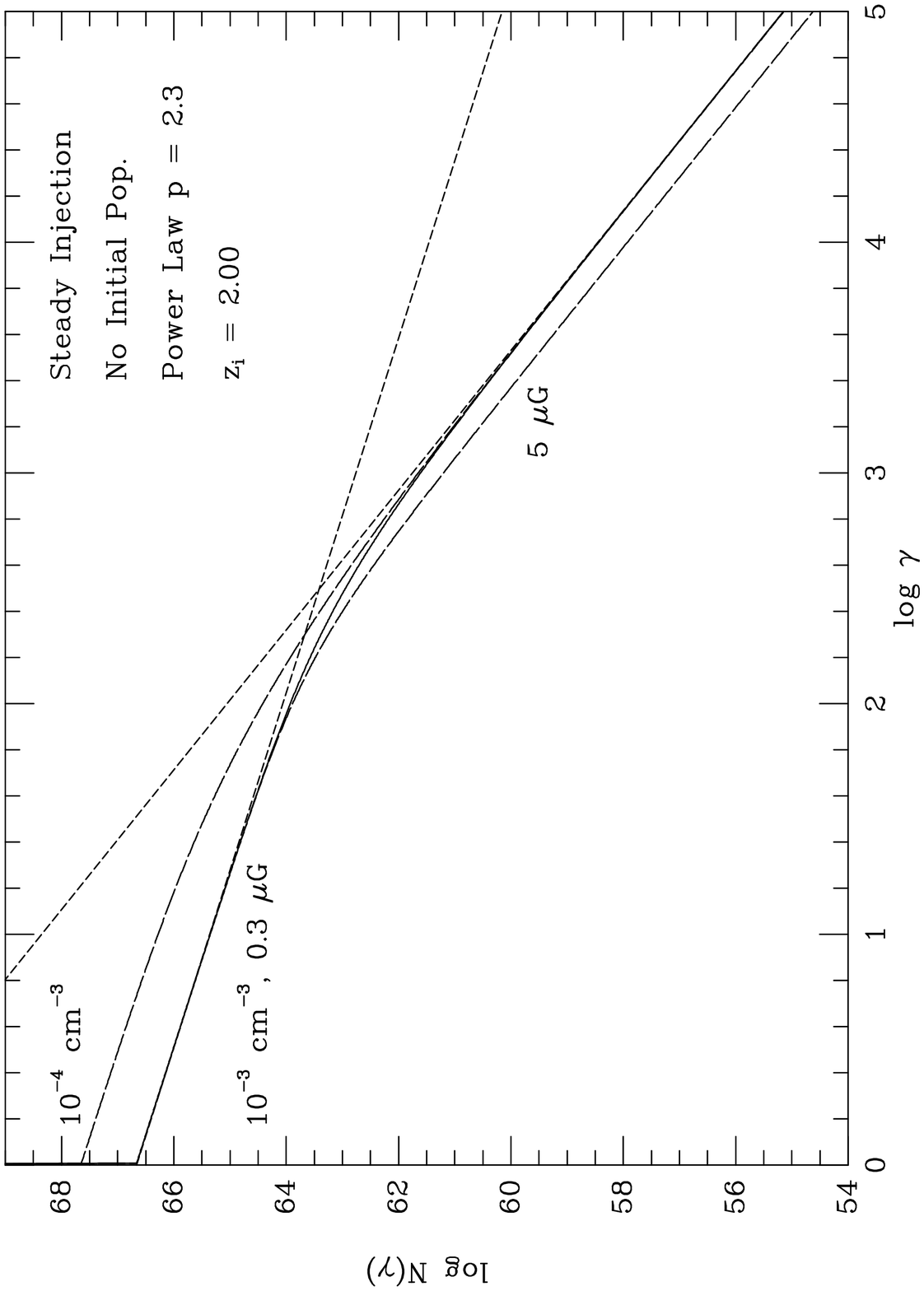}
\figcaption{
The relativistic electron population in a cluster in which particles
have been injected at a constant rate since $z_i = 2$.
The injected particles had a power-law distribution with $p = 2.3$.
The solid curve (Model 1) is the distribution in a cluster with an average
thermal electron density of $n_e = 10^{-3}$ cm$^{-3}$.
The two short-dashed lines are power-law distributions with $p^\prime = 1.3$ and
3.3, which match the resulting particle distributions at low and high
energy, respectively.
The two long-dashed curves are similar to Model 1, but have a decreased
electron density of $n_e = 10^{-4}$ cm$^{-3}$ (Model 7)
or an increased magnetic field of $B = 5$ $\mu$G (Model 8).
\label{fig:model_steady}}

\vskip0.2truein

\subsection{Solutions} \label{sec:numerical_solutions}

% \placetable{tab:results}

Some of the results for the numerical models for the electron population 
in clusters are given in Table~\ref{tab:results}.
For each model, the second column gives the total number of relativistic
electrons $N_{tot}$ at present (at $z = 0$).
Columns 3 and 4 give the kinetic energy in relativistic electrons at
present ($z = 0$), either for all of the electrons ($\gamma > 1$)
or only for those with $\gamma > 300$.
The fifth column gives the energy-averaged mean value of $\gamma$,
defined here as
\begin{equation} \label{eq:gmean}
\langle \gamma \rangle \equiv
\frac{\int N(\gamma) \gamma^2 \, d \gamma}
{\int N(\gamma) \gamma \, d \gamma}
\, .
\end{equation}
Sarazin \& Lieu (1998) show that this quantity is useful for estimating
the emission properties of  these electrons.
For models without particle injection at the present time
(\S~\ref{sec:numerical_solutions_initial}), the last column gives the upper
cut off to the electron distribution, $\gamma_{max}$.

\subsubsection{Steady Particle Injection}
\label{sec:numerical_solutions_steady}

I first considered models in which relativistic electrons are injected
into the ICM at a constant rate over the history of the cluster, and
there was no initial population of electrons.
Figure~\ref{fig:model_steady} shows the result of steady injection since
an initial redshift of $z_i = 2$;
the solid curve is for Model 1 in Table~\ref{tab:models}.
For this large redshift, the age of the cluster exceeds the loss time
of the electrons for all values of $\gamma$ (Figure~\ref{fig:lifetime}).
This is particularly true when one includes the increase in the energy
density of the CMB with redshift.
For example, Figure~\ref{fig:gmax} shows that IC losses will remove
all of the particles produced at $z \sim 1$ down to low values of
$\gamma$ where Coulomb losses dominate and the lifetime is short.
Thus, one expects that the particle distribution will approach steady
state, with the rate of injection balancing the rate of loss at most
energies.
The injected electrons had a simple power-law distribution
(equation~\ref{eq:injection_pw1}) with $p = 2.3$.
Figure~\ref{fig:model_steady} shows that the resulting electron population
is essentially a broken power-law.
The dashed lines show power-law distributions with indices of
$p^\prime = 1.3$ and 3.3.
At lower energies where Coulomb losses dominate ($\gamma \la 100$),
the distribution is very nearly a power-law with $p^\prime = 1.3$.
This is one power flatter than the injected population, as expected from
the steady-state result of equation~(\ref{eq:steady_low}).
At high energies where IC and synchrotron losses dominate
($\gamma \ga 1000$), the distribution is nearly a power-law with
$p^\prime = 3.3$.
This is one power steeper than the injected population, as expected from
the steady-state result of equation~(\ref{eq:steady_high}).

%
% table of model results
%
% SUBMISSION
% \begin{table}[thb]
% \caption{\hfil Results for the Electron Energy Spectrum \label{tab:results} \hfil}
% PREPRINT
% \begin{table*}[htb]
% \caption{\hfil Results for the Electron Energy Spectrum \label{tab:results} \hfil}
{
\small
\begin{center}
\tabcaption{\hfil Results for the Electron Energy Spectrum \label{tab:results} \hfil}
\begin{tabular}{lccccc}
\tableline
\tableline
&&\multicolumn{2}{c}{$E_{CR,e}$}&$\langle \gamma \rangle$&$\gamma_{max}$\cr
&$N_{tot}$&$\gamma > 1$&$\gamma > 300$&&\cr
Model&($10^{66}$)&\multicolumn{2}{c}{($10^{61}$ ergs)}&&\cr
\tableline
 1&   \phn11.98&   \phn28.39&\phn9.48&      \phn544.6&\cr
 2&   \phn14.96&   \phn35.46&   11.85&      \phn544.6&\cr
 3&   \phn21.16&   \phn50.15&   16.81&      \phn546.2&\cr
 4&   \phn29.12&   \phn67.32&   23.50&      \phn564.2&\cr
 5&   \phn65.38&      119.23&   42.59&      \phn675.7&\cr
 6&      377.34&      264.52&   71.55&         1159.4&\cr
 7&   \phn95.67&   \phn91.58&   11.50&      \phn204.7&\cr
 8&   \phn11.73&   \phn21.10&\phn4.29&      \phn302.0&\cr
 9&      118.87&      198.19&   62.98&      \phn761.9&         15107\cr
10&\phn\phn7.67&   \phn56.76&   23.17&      \phn334.0&      \phn1379\cr
11&\phn\phn1.95&   \phn13.38&\phn0.70&      \phn153.6&   \phn\phn338\cr
12&\phn\phn0.67&\phn\phn2.17&\phn0.00&   \phn\phn60.6&   \phn\phn110\cr
13&\phn\phn0.00&\phn\phn0.00&\phn0.00&\phn\phn\phn3.1&\phn\phn\phn83\cr
14&   \phn44.38&   \phn88.65&\phn4.17&   \phn\phn93.2&   \phn\phn405\cr
15&\phn\phn3.93&   \phn28.99&\phn7.67&      \phn209.9&   \phn\phn582\cr
16&\phn\phn1.95&   \phn13.43&\phn0.74&      \phn154.4&   \phn\phn340\cr
17&\phn\phn1.93&   \phn12.82&\phn0.31&      \phn147.0&   \phn\phn320\cr
18&\phn\phn1.76&\phn\phn9.01&\phn0.00&      \phn104.7&   \phn\phn211\cr
19&\phn\phn1.44&\phn\phn4.66&\phn0.00&   \phn\phn61.6&   \phn\phn114\cr
20&\phn\phn8.51&   \phn21.01&\phn6.93&      \phn527.5&\cr
21&\phn\phn0.54&   \phn11.12&\phn8.13&         1513.9&\cr
22&\phn\phn0.61&\phn\phn7.57&\phn0.93&      \phn212.0&   \phn\phn333\cr
23&      113.10&      159.80&   53.50&      \phn787.0&\cr
24&   \phn57.52&   \phn86.59&   27.10&      \phn740.1&\cr
25&   \phn29.74&   \phn49.98&   13.90&      \phn664.4&\cr
26&   \phn13.06&   \phn28.02&\phn5.98&      \phn521.8&\cr
27&\phn\phn3.06&   \phn14.84&\phn1.23&      \phn224.6&\cr
28&\phn\phn2.06&   \phn13.52&\phn0.75&      \phn161.5&\cr
\tableline
\end{tabular}
\end{center}
% SUBMISSION
% \end{table}
% PREPRINT
% \end{table*}
}

\vskip0.2truein

The approach to steady-state in models with constant injection
is shown in Figure~\ref{fig:model_steady_z}, which gives
the present day populations in models with initial redshifts of
$z_i = 0.01$, 0.1, 0.3, 0.5, 1, and 2 (top to bottom).
(These are Models 1--6 in Table~\protect\ref{tab:models}.)
All of the models have an average electron density of $n_e = 10^{-3}$
cm$^{-3}$, and a magnetic field of $B = 0.3$ $\mu$G.
The particles are injected with a power-law spectrum with $p = 2.3$.
The older models ($z_i = 0.5$, 1, and 2) have achieved steady-state, since
the loss times of electrons of all energies are less than the age of
the cluster.
The younger cluster models show departures from steady-state.
For these models, the energy spectra of electrons can be divided into
three regions.
At very high energies ($\gamma \ga \gamma_{max}$),
the IC losses are sufficiently rapid that the loss time scale
(eqn.~\ref{eq:life_instant}) is shorter than
the age, and the populations approach steady-state, and the population is
a power-law which is one power steeper than the injected spectrum.
Similarly, at low energies the loss time scale is also generally shorter
than the age (Figure~\ref{fig:lifetime}).
The population also approaches steady-state below some lower value
of $\gamma \la \gamma_{low}$, and population is a power-law which is one
power flatter than the injection spectrum.
At intermediate energies $\gamma_{low} \ll \gamma \ll \gamma_{max}$,
the loss time scale is longer than the age, and the population is
given by the accumulation of the injection rate.
Thus, it has the same spectrum as the injection spectrum.
For younger injection models where $\gamma_{low} \ll \gamma_{max}$, the
resulting electron population is given quite accurately by the
self-similar solution (equation~\ref{eq:self_similar}).

\centerline{\null}
\vskip2.55truein
\includegraphics{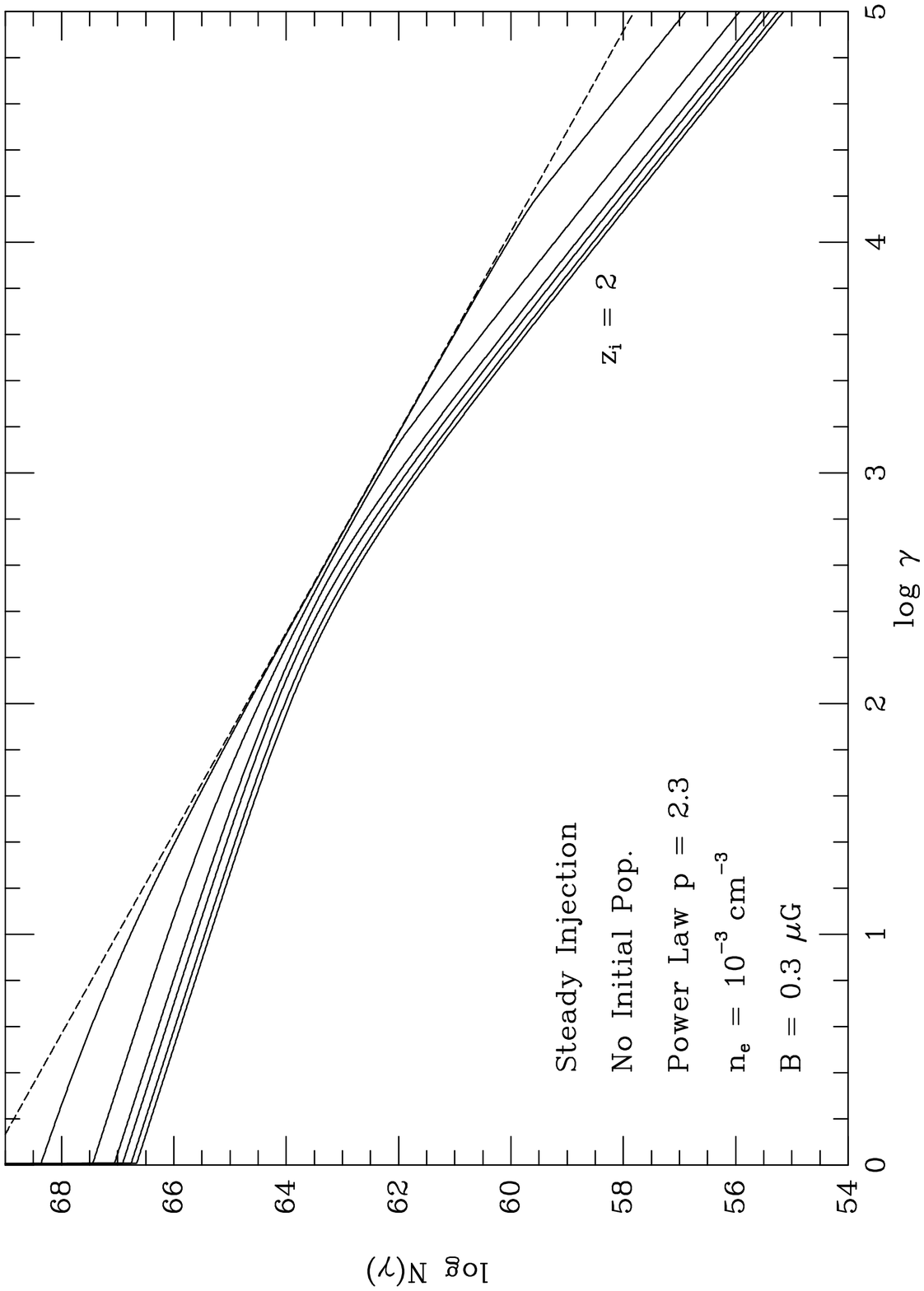}
\figcaption{
The present day relativistic electron populations in a series of models with 
steady particle injection but no initial population of particles.
The solid curves show models for clusters which started at redshifts of
$z_i = 2$, 1, 0.5, 0.3, 0.1, and 0.01 (bottom to top).
These are Models 1--6 in Table~\protect\ref{tab:models}.
In all of the models, the particles are injected at a constant rate with
a power-law distribution with $p = 2.3$.
The short-dashed curve gives the total power-law spectrum of all of the
injected particle over the cluster lifetime.
\label{fig:model_steady_z}}

\vskip0.2truein

Figure~\ref{fig:model_steady} also shows the effect of changing the
gas density or magnetic field in the models.
The two long-dash curves are the same as Model 1, but with the
electron density decreased to $n_e = 10^{-4}$ cm$^{-3}$ (Model 7)
or the magnetic field increased to $B = 5$ $\mu$G (Model 8).
Increasing or decreasing the gas density mainly increases or decreases
the rate of Coulomb losses at low energies, and this decreases or
increases the electron population at low energies.
Reducing $n_e$ also decreases the electron energy or value of $\gamma$
at which the transition between the two power-law slopes occurs in
steady-state.
Similarly, increasing magnetic field enhances the rate of synchrotron
losses at high energies.
This reduces the population of at high energies, but the effect is only
important if the field exceeds about 3 $\mu$G.

\subsubsection{Initial Population with No Later Particle Injection}
\label{sec:numerical_solutions_initial}

Next, models were calculated with an initial electron population at
a redshift $z_i$, but with no subsequent injection of particles.
Figure~\ref{fig:model_init_z} shows the result for models with differing
values of the initial redshift $z_i$.
The initial particle population was a single power-law with
$p_0 = 2.3$
(eqn.~\ref{eq:initial_pw1})
At high energies, the electron population is reduced and the shape of
the spectrum steepened by IC and synchrotron losses.
There is an upper cut-off to the electron distribution
(eqns.~\ref{eq:distr_IC_nocosmo}--\ref{eq:distp_IC_nocosmo}) for
values of $\gamma$ beyond a maximum values $\gamma_{max}$,
given approximately by equation~(\ref{eq:gmax_IC_cosmo}), or
equation~(\ref{eq:gmax_IC_nocosmo}) for low redshifts.
The cutoff energies in the numerical models are in reasonable agreement with
these analytic approximations as long as $\gamma_{max} \ga 300$.
The value of $\gamma_{max}$ decreases as the age of the
electron populations (or, equivalently, the initial redshift $z_i$)
increases.

\centerline{\null}
\vskip2.55truein
% \special{psfile=model_init_z.ps angle=-90 voffset=+190 hoffset=-19
\includegraphics{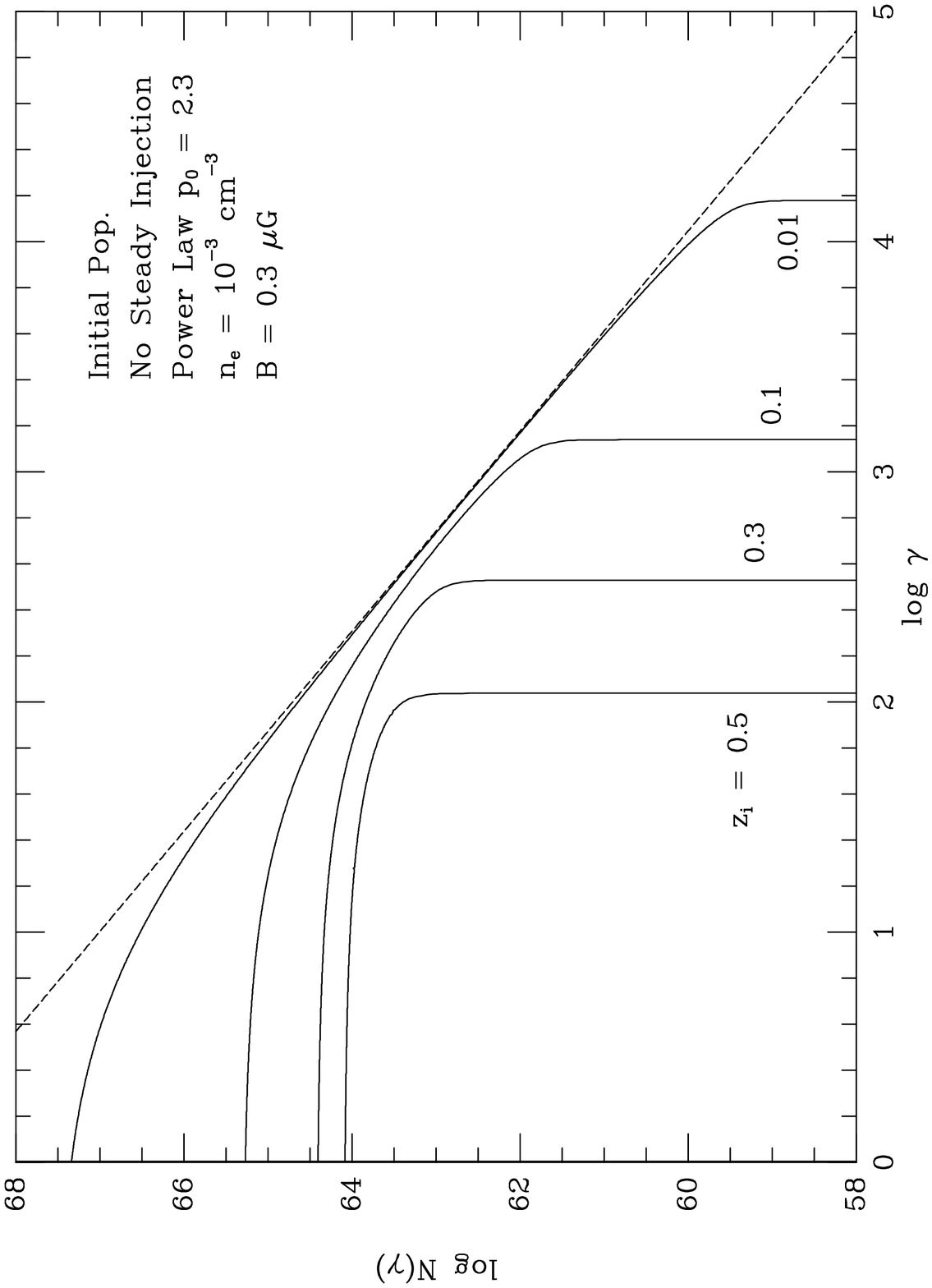}
\figcaption{
The relativistic electron population in models with an initial population
of particles generated at a redshift $z_i$, but no subsequent injection.
The initial population, which is shown as a dashed line, had a power-law
distribution with $p_0 = 2.3$.
The solid curve show the resulting present population for values of
the initial redshift of $z_i = 0.01$, 0.1, 0.3, and 0.5.
\label{fig:model_init_z}}

\vskip0.2truein

In order to have any significant population of primary electrons with
$\gamma \ga 10^2$ at the present time,
Figures~\ref{fig:gmax} and \ref{fig:model_init_z} show that
there must have been a substantial injection of particles into clusters
at moderately low redshifts, $z \la 1$.

\centerline{\null}
\vskip2.55truein
\includegraphics{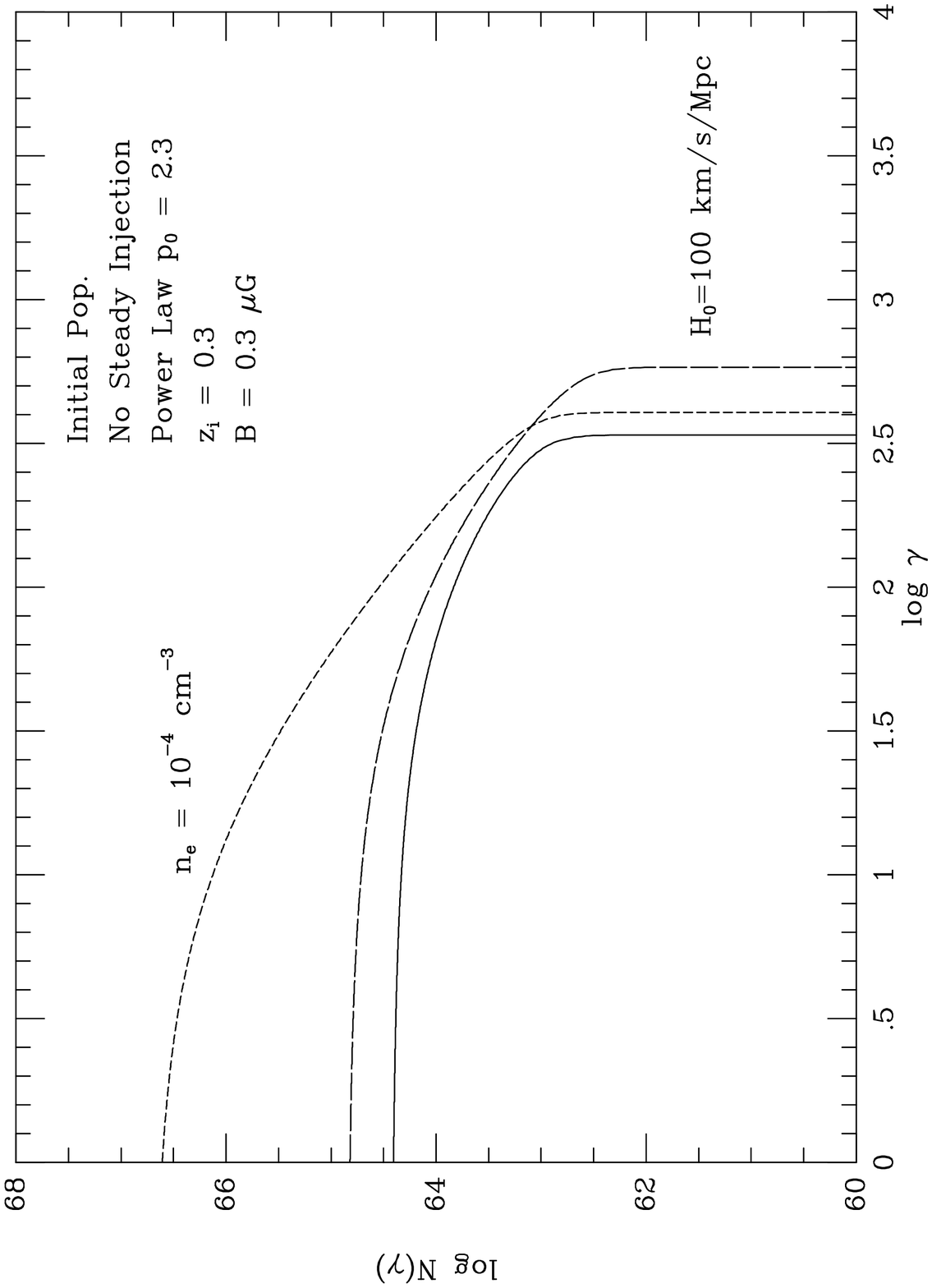}
\figcaption{
The effect on the electron population of changing several of the parameters.
The solid curve is Model~11 for an initial population
of particles generated at a redshift $z_i = 0.3$, but no subsequent injection.
The short-dashed curve (Model~14) shows the result of reducing the average
electron density from $10^{-3}$ to $10^{-4}$ cm$^{-3}$.
The long-dashed curve (Model~15) shows the result of increasing the Hubble
constant from 65 to 100 km s$^{-1}$ Mpc$^{-1}$.
\label{fig:model_init2}}

\vskip0.2truein

At low energies ($\gamma \la \gamma_{low}$), the electron population
is reduced and the shape of the distribution flattened by Coulomb losses
to the thermal plasma.
As discussed in \S~\ref{sec:analytic_init_low}, the slow variation
of the Coulomb loss function with $\gamma$ results in a flat energy
spectrum at low energies (eqn.~\ref{eq:low_energy_soln}).
All of the older cluster models in Figure~\ref{fig:model_init_z} show
this behavior.
In the younger models ($z_i \la 0.3$), there is an extended region
$\gamma_{low} \la \gamma \la \gamma_{max}$ where the initial population
is preserved.
In the older models ($z_i \ga 0.3$), the flat energy spectrum at low
energies and steep cut-off at high energies nearly meet, and the
energy spectra have a ``top hat'' form.

Figure~\ref{fig:model_init2} shows the effect of varying two of the
basic parameters of the models.
The solid curve is Model~11, with an average electron density of
$n_e = 10^{-3}$ cm$^{-3}$ and a Hubble constant of
$H_0 = 65$ km s$^{-1}$ Mpc$^{-1}$.
In all of these models, the electron population was injected into the
cluster at $z_i = 0.3$.
The short-dashed curve (Model~14) is a model with all of the same parameters,
but with a lower value for the electron density of $n_e = 10^{-4}$
cm$^{-3}$.
Reducing the electron density lowers the rate of Coulomb losses,
which dominate at low energies.
As a result, increasing or decreasing the electron density decreases
or increases the population of low energy electrons.

Changing the value of the Hubble constant changes the age of the cluster
for a fixed value of $z_i$ and $q_0$.
The long-dashed curve in Figure~\ref{fig:model_init2} shows Model~15, which
has all of the same parameter, except the Hubble constant is increased
to $H_0 = 100$ km s$^{-1}$ Mpc$^{-1}$.
Increasing the Hubble constant shortens the age, and reduces the effect
of losses.
Thus, the electron population is larger at all energies, and the cutoff
value of $\gamma_{max}$ is increased.
Lowering the Hubble constant has the opposite effect.

\centerline{\null}
\vskip2.55truein
\includegraphics{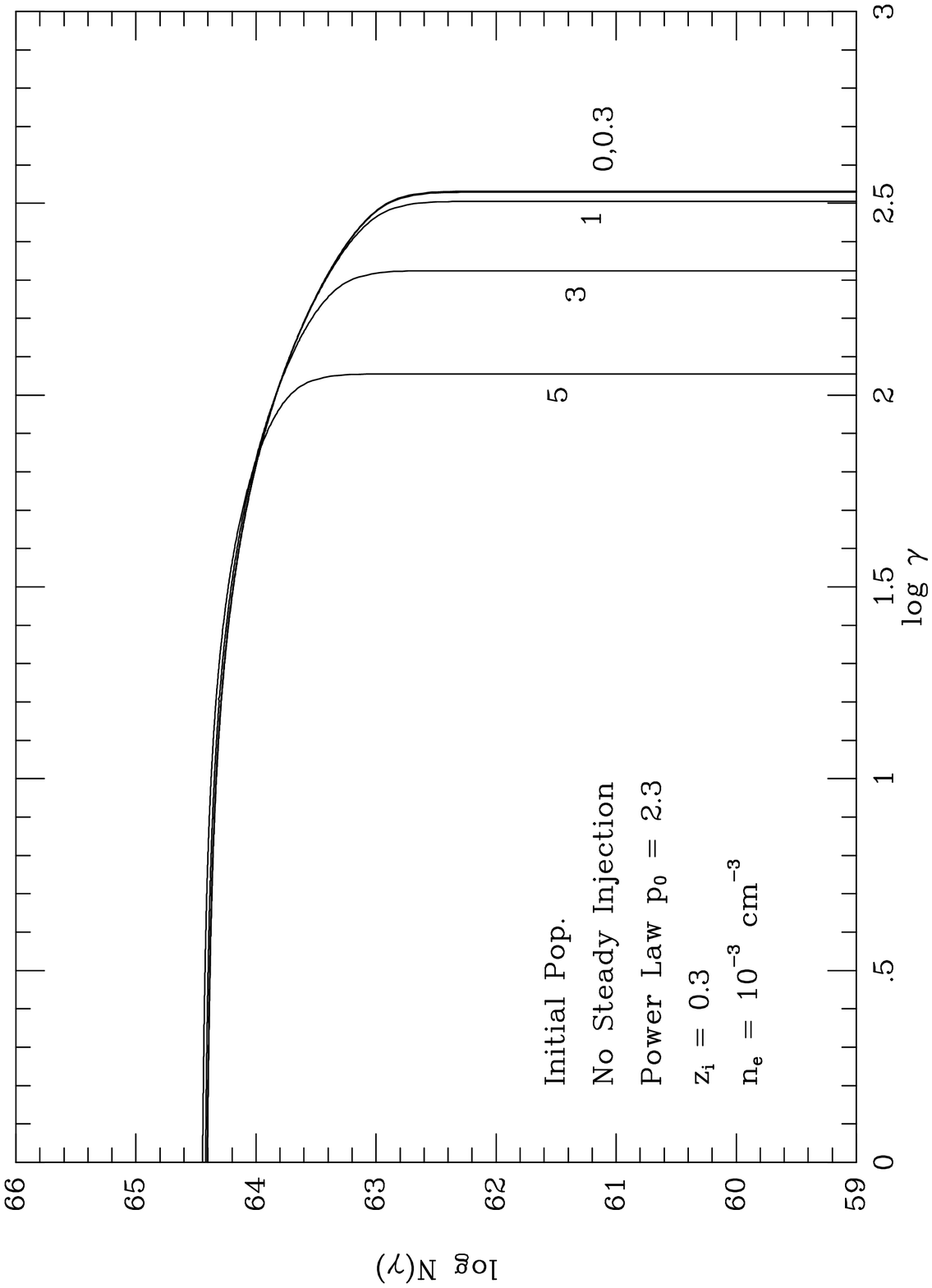}
\figcaption{
Models for relativistic electrons with a variety of
intracluster magnetic field strengths (Models~11, 16--19).
The curves show the populations for magnetic fields of
0, 0.3, 1, 3, and 5 $\mu$G.
The curves for no magnetic field and 0.3 $\mu$G are nearly
indistinguishable.
\label{fig:model_B}}

\vskip0.2truein

I also constructed models with a variety of average values for the
intracluster magnetic field, which can affect the electron population
through synchrotron losses.
Figure~\ref{fig:model_B} shows Models 16, 11, and 17--19, with magnetic
fields of $B = 0$, 0.3, 1.0 3.0, and 5.0 $\mu$G.
All of the other parameters of Models 16--19 are identical to Model~11.
Because synchrotron losses have the same variation with $\gamma$
as IC losses, they affect the electron population in a very similar
way.
Unless the ICM magnetic field exceeds 1 $\mu$G, synchrotron losses
do not have a very significant effect.
For larger values of $B$, the number of high energy electrons and
the value of $\gamma_{max}$ are reduced as $B$ is increased.
The values of $\gamma_{max}$ are in good agreement with those predicted
by equation~(\ref{eq:gmax_B_cosmo}).
However, to produce an appreciable effect, the magnetic field is required
to be considerably larger than expected in clusters
(Kim et al.\ 1990, 1991:
Kronberg 1994;
Fusco-Femiano et al.\ 1998).

\centerline{\null}
\vskip2.55truein
\includegraphics{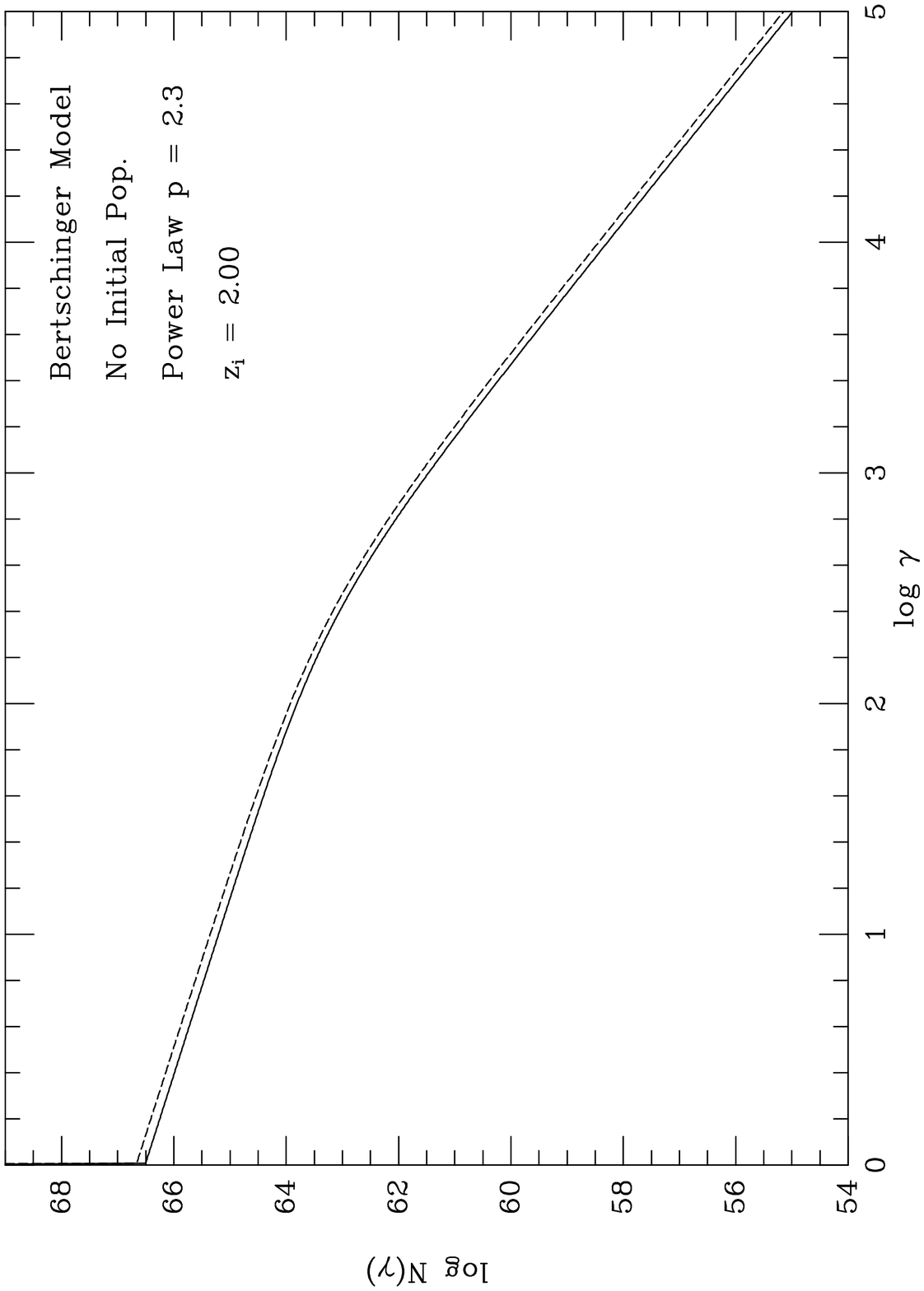}
\figcaption{
The solid curve is the
relativistic electron population in a model based on the Bertschinger
(1985) self-similar solution for secondary infall in a cluster
(eq.~[\protect\ref{eq:injection_bert}], Model~20).
For comparison, the dashed curve is the steady injection model with
the same properties (Model~1).
\label{fig:model_bert}}

\vskip0.2truein

\subsubsection{Self-Similar Cluster Accretion Shock}
\label{sec:numerical_solutions_bert}

I also calculated a model in which particles were injected with
a rate proportional by the rate of deposition of shock energy
(eq.~\ref{eq:injection_bert})
in the self-similar secondary infall model of
Bertschinger (1985).
In Figure~\ref{fig:model_bert}, this model (Model~20) is compared to
the steady injection model (Model~1) in which the rate of injection
of particles is a constant.
The two models give very similar results.
The small differences between the self-similar secondary infall model
and the steady injection model are mainly due to differences in the
normalization
(\S~\ref{sec:source_normalization}).
The models are normalized based on the total input of relativistic
electrons, and the Bertschinger model and the steady model have different
variations in the rate of injection with time or redshift
(eq.~\ref{eq:injection_bert}).
If the present day rate of injection is the same, the maximum difference
between the Bertschinger model and the steady injection model is only about
10\%, and occurs near the knee in the particle spectrum (at $\gamma \approx
160$ in Figure~\ref{fig:model_bert}).
The very near agreement of the Bertschinger model and the steady injection
model reflects the fact that both models have nearly reached steady-state,
and the current electron spectrum is mainly determined by the present rate
of injection, rather that the past rate over cosmological time scales.

\subsubsection{Models with Broken Power Law Particle Injection}
\label{sec:numerical_solutions_break}

In \S~\ref{sec:source_spectrum}, we noted that the spectrum of cosmic
ray protons produced by shock acceleration is expected to flatten
somewhat at kinetic energies $E \la 1$ GeV, where the protons
become nonrelativistic.
In most models, it is expected that the electron spectrum will
remain nearly a power-law to much lower energies.
Here, we consider that possibility that the spectrum of electrons
also flattens below 1 GeV.
I adapt the simple two power-law form of equations~\ref{eq:initial_pw2})
and (\ref{eq:injection_pw2}).
In Figure~\ref{fig:model_break}, the resulting electron populations 
in models with broken power-law models for the initial population or
for the injection spectrum.
For models with steady injection (Model~21), the broken spectrum produces
a low energy electron spectrum which is a flatter power-law by the same
amount that the injection spectrum if flattened at low energies.
For models with initial broken power-law spectra, the population is
very flat at low energies, and has the same high energy cut-off as the
model with a single power-law.
However, the broken power-law models have a peak at energies just below
$\gamma_{max}$ if $\gamma_{br} > \gamma_{max}$.

\centerline{\null}
\vskip2.55truein
\includegraphics{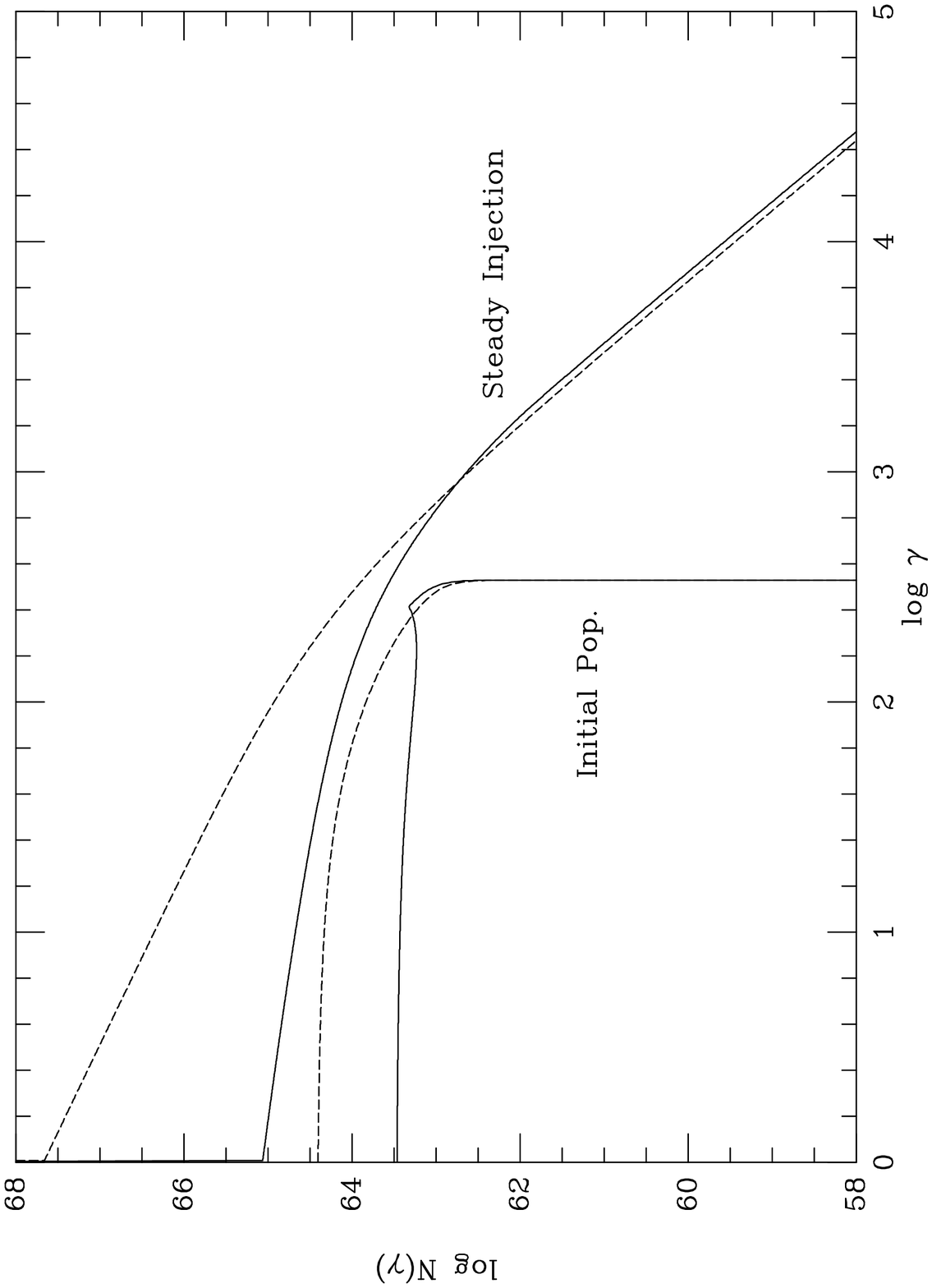}
\figcaption{
Models with broken power-law injection (eq.~\protect\ref{eq:injection_pw2})
or initial spectra (eq.~\protect\ref{eq:initial_pw2}) are compared to models
with simple power-laws.
The lower curves compare a model with an initial population 
(at $z_i = 0.3$ ) which is a broken power-law (solid curve, Model~22)
with an otherwise identical model with a single power-law
(dashed curve, Model 11).
In these models, there is no subsequent particle injection.
The upper curves compare models with no initial population but
a steady rate of particle injection since $z_i = 2$.
The solid curve has a broken power-law injection spectrum (Model 21),
while the dashed curve is for a single power-law (Model 1).
For clarity of presentation, the steady injection curves were displaced
upwards by one order of magnitude.
\label{fig:model_break}}

% \vskip0.2truein

\subsubsection{Models with Both an Initial Population and Steady Injection}
\label{sec:numerical_solutions_both}

Finally, we consider models in which there is an initial population in
the cluster at $z_i$, and in which there also is current injection of
new particles.
I assume that this current injection has occurred at a constant rate since
redshift $z_s$.
These two components might represent particle populations which were produced
by shocks associated with the formation of the cluster at $z_i$ and by
a current merger event, which started at $z_s$.
Because the equation for the evolution of the electron population is linear
(eq.~\ref{eq:evolution}), these models are just a linear combination of
the models discussed in \S\S~\ref{sec:numerical_solutions_steady}
and \ref{sec:numerical_solutions_initial}.
An illustrative set of such models are shown in Figure~\ref{fig:model_both}.
I assumed $z_i = 0.3$, $z_s = 0.05$, $n_e = 10^{-3}$ cm$^{-3}$, and
$B = 0.3$ $\mu$G in these models.
Both the initial population and current injection are given by a single
power-law spectrum with $p = p_0 = 2.3$.
The different models are characterized by differing values of $F_{inj}$,
which is the fraction of the particle energy which has been injected by
the current steady injection process, as opposed to the initial
population.
Models are shown with values of
$F_{inj} = 100$\%, 50\%, 25\%, 10\%, 1\%, 0.1\%, and 0\%.
(Models 23--28 and 11).

For models with intermediate values of the energy fraction due to the
ongoing injection, the energy spectrum at low energies is flatter than
the $p^\prime = 1.3$ power-law due to steady state injection, but steeper
then the flat distribution expected for the initial population.
There is a rapid drop-off in the electron population at $\gamma_{max}$.
Above $\gamma_{max}$, the electrons are only due to current injection.
Of course, in a real cluster it is unlikely that the initial population
would have been produced at a single redshift.
Thus, the fall off at $\gamma_{max}$ is likely to be more gradual, and
there might be a more continuous transition to the current injection
population above $\gamma_{max}$.
However, these models should give a general sense of the electron
energy spectra to be expected in real clusters.

\centerline{\null}
\vskip2.55truein
\includegraphics{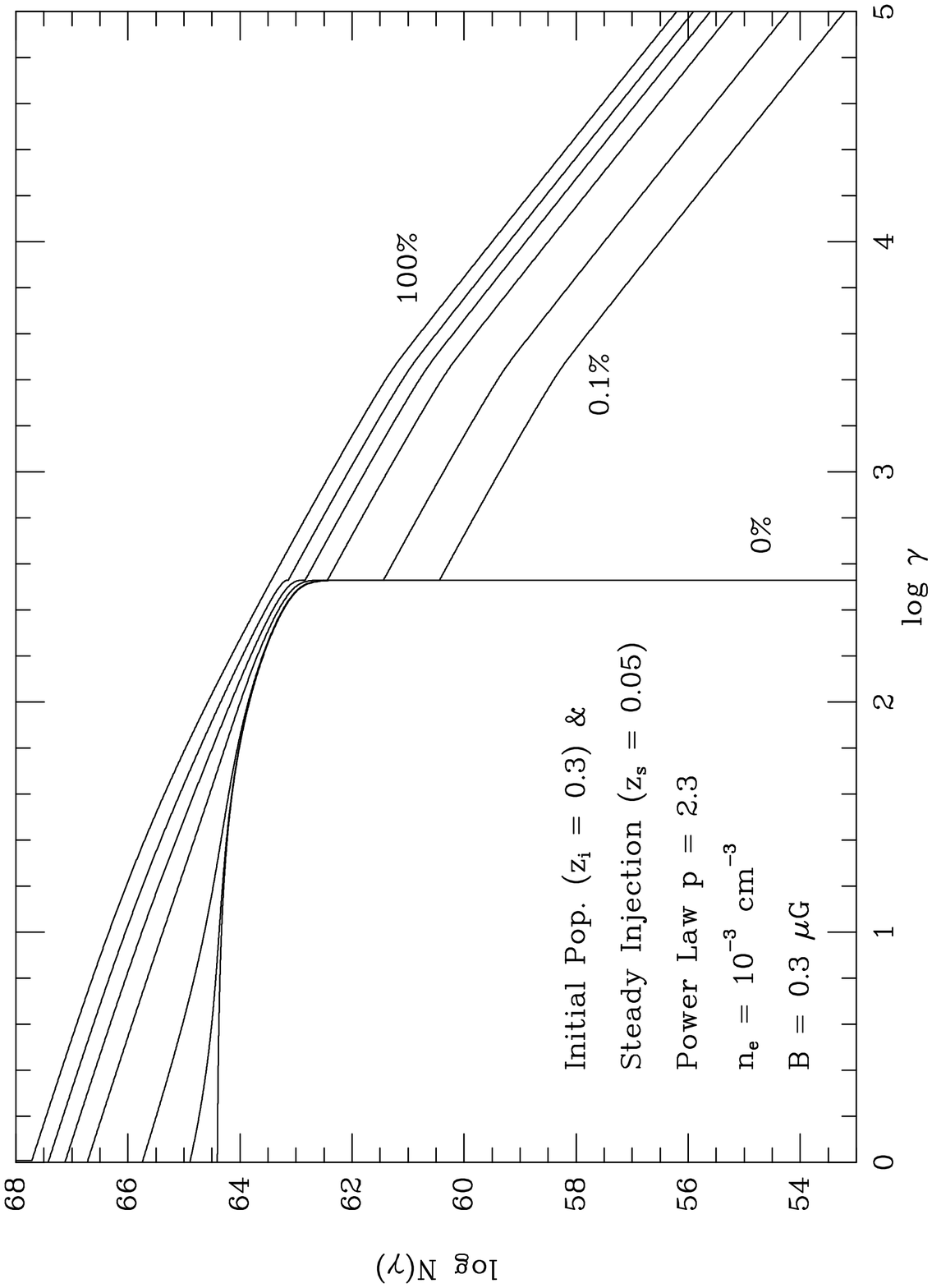}
\figcaption{
The electron energy spectra in models with both an initial electron
population starting at $z_i = 0.3$ and continual injection of new particles
since $z_s = 0.05$.
All of the models have a single power-law spectrum for particle injection
with $p = p_0 = 2.3$, an electron density of $n_e = 10^{-3}$ cm$^{-3}$,
and a magnetic field of $B = 0.3$ $\mu$G.
The values of the fraction of particle energy due to current injection
are (top to bottom)
$F_{inj} = 100$\%, 50\%, 25\%, 10\%, 1\%, 0.1\%, and 0\%.
\label{fig:model_both}}

% \vskip0.2truein

\section{Inverse Compton Emission} \label{sec:ic}

I also calculated the emission produced by the relativistic electron
populations due to the inverse Compton scattering of CMB photons.
Let $L_\nu d \nu$ be the luminosity of IC emission at frequencies from
$\nu$ to $\nu + d \nu$.
Then, the spectrum of IC emission is related to the electron spectrum
$ N ( \gamma )$ by
\begin{equation} \label{eq:ic}
L_\nu = 12 \pi \sigma_T \,
\int_1^\infty N ( \gamma ) \, d \gamma \,
\int_0^1 J \left( \frac{\nu}{4 \gamma^2 x } \right) \, {\cal F} ( x ) \, dx \, ,
\end{equation}
where
\begin{equation} \label{eq:fic}
{\cal F} ( x ) \equiv 1 + x + 2 x \ln x - 2 x^2 \, .
\end{equation}
Here, $J ( \nu )$ is the mean intensity at a frequency $\nu$ of the
radiation field being scattered.
For the CMB, this is just the black body function
\begin{equation} \label{eq:bb}
J ( \nu ) = B_\nu ( T_{CMB} ) = \frac{2 h \nu^3}{c^2} \,
\frac{1}{\exp ( h \nu / k T_{CMB} ) - 1} \, .
\end{equation}

Figure~\ref{fig:ic_steady} shows the IC spectra of a series of models
with a steady rate of particle injection, but no initial electron
population.
These are the same models (Models~1--6) whose electron populations
were shown in Figure~\ref{fig:model_steady_z}.
The models shown have had steady injection starting at redshifts
$z_i = 2$, 1, 0.5, 0.3, 0.1, and 0.01 (bottom to top) and continuing to
the present.
The dashed line is the spectrum produced by a model in which the electron
population is just that injected into the cluster;
since this involves no evolution, this model is equivalent to $z_i = 0$.
For this model with no evolution, the spectrum is a power-law
$L_\nu \propto \nu^\alpha$, where $\alpha = - ( p - 1 ) / 2$.
This gives $\alpha = -0.65$ for the injection spectrum of $p = 2.3$.
At frequencies above $10^{17}$ Hz, the spectrum quickly steepens with time
due to IC losses, and becomes a half power steeper,
$\alpha \approx - ( p / 2 ) \approx -1.15$.
At low frequencies, the spectrum flattens more gradually due to Coulomb
losses, approaching a spectrum which is a half power flatter,
$\alpha \approx - ( p - 2 ) / 2 \approx -0.15$.
If the steady injection has occurred since a redshift of $ z_i \ga 0.2$,
the spectrum is near that of a steady-state electron population, with
these two power-laws meeting at a knee at $\nu \sim 3 \times 10^{16}$ Hz.

\centerline{\null}
\vskip2.55truein
\includegraphics{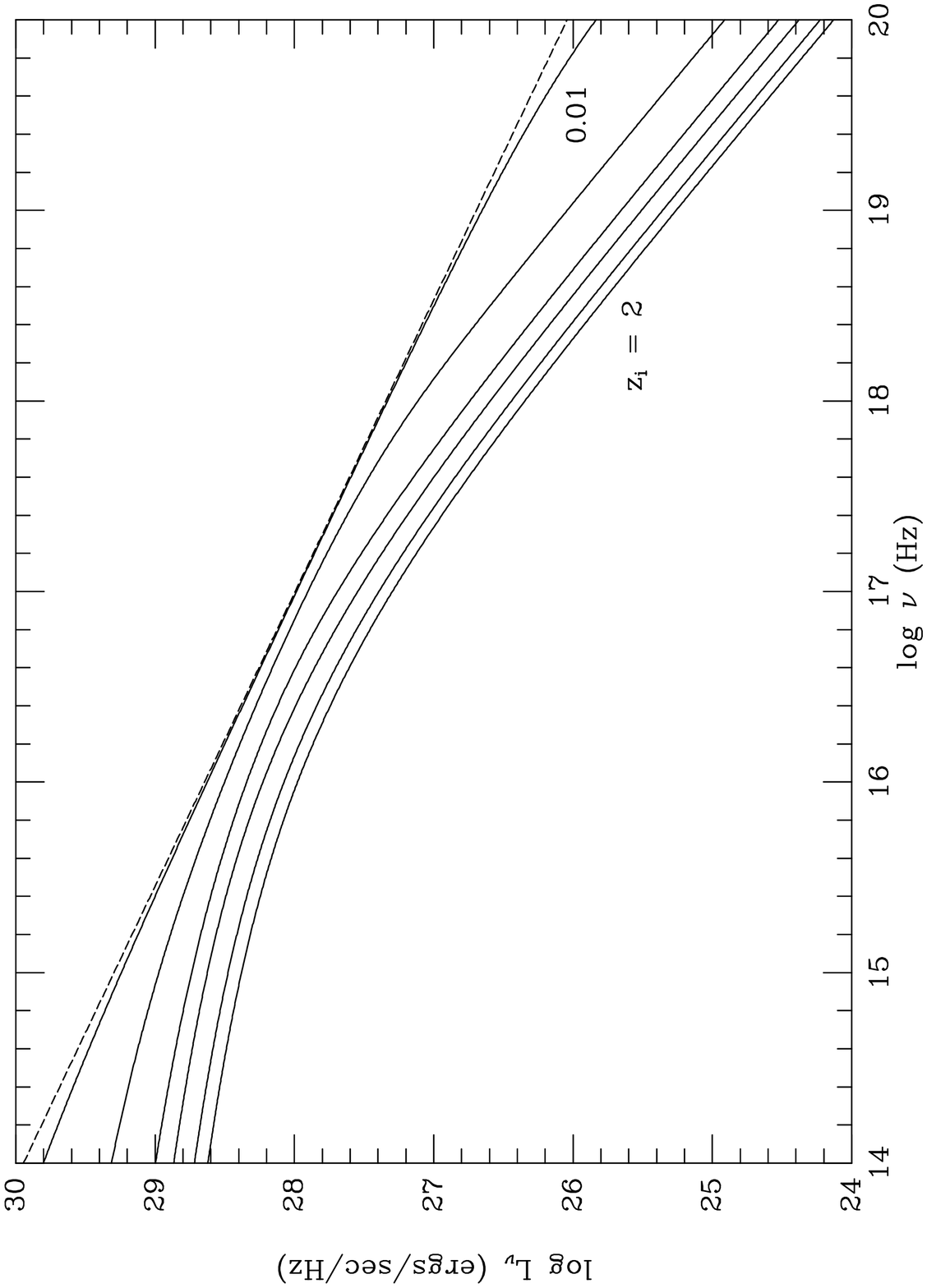}
\figcaption{
The IC emission spectra for cluster models
with steady particle injection but no initial population of particles.
The notation is similar to that in
Figure~\protect\ref{fig:model_steady_z}.
The solid curves show spectra for clusters which started at redshifts of
$z_i = 2$, 1, 0.5, 0.3, 0.1, and 0.01 (bottom to top).
These are Models 1--6 in Table~\protect\ref{tab:models}.
The dashed curve gives the spectrum for $z_i = 0$ (i.e., no evolution
of the electron population due to losses).
\label{fig:ic_steady}}

% \vskip0.2truein

\centerline{\null}
\vskip2.55truein
\includegraphics{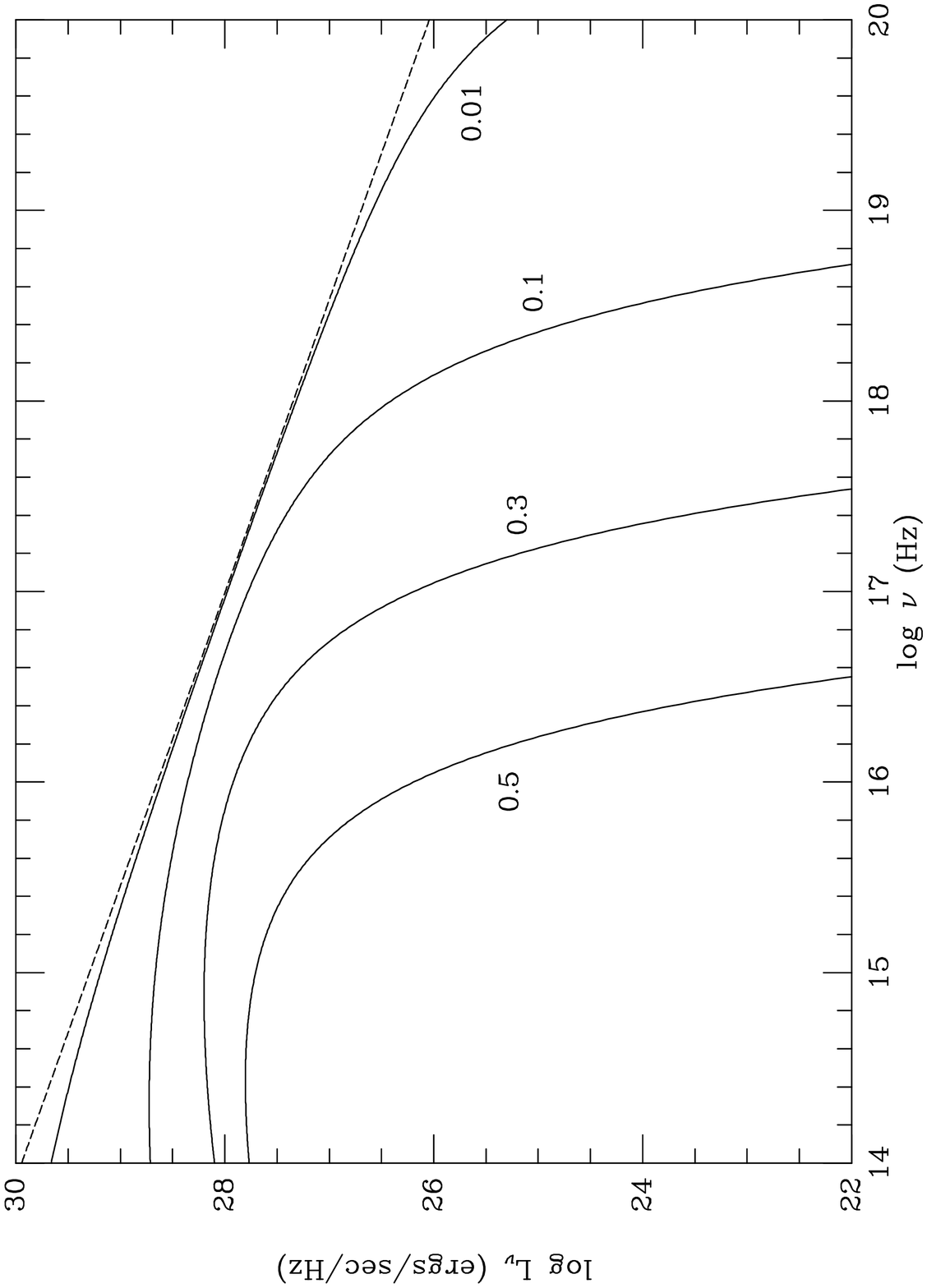}
\figcaption{
The IC emission spectra for cluster models
with an initial population of electrons but no subsequent
injection.
The notation is similar to that in
Figure~\protect\ref{fig:model_init_z}.
The solid curves show spectra for clusters which started at redshifts of
$z_i = 0.01$, 0.1, 0.3, and 0.5.
These are Models 9--12 in Table~\protect\ref{tab:models}.
The dashed curve gives the spectrum for $z_i = 0$ (i.e., no evolution
of the electron population due to losses).
\label{fig:ic_initial}}

\vskip0.2truein

The IC spectra of models with an initial electron population but no
subsequent acceleration are shown in Figure~\ref{fig:ic_initial}.
This Figure gives the spectra for the same models shown in
Figure~\ref{fig:model_init_z}, in which the initial population of
electrons occurs at redshifts of
$z_i = 0$, 0.01, 0.1, 0.3, and 0.5.
The electron energy distributions of these models have a cut off at high
energies, $\gamma_{max}$, which results from
rapid IC and synchrotron losses by high energy electrons
(see eqs.~\ref{eq:gmax_IC_nocosmo}, \ref{eq:gmax_IC_cosmo},
\ref{eq:gmax_B_cosmo}, and Figure~\ref{fig:gmax}).
As a result, the IC spectra have a very rapid fall off at high
energies.
If the initial electron spectrum is a power-law with an index
of $p_0$ (eq.~\ref{eq:initial_pw1}), and if $\gamma_{max}$ is large enough
that the energy losses are dominated by IC and synchrotron losses, then
the electron energy spectrum at a later time is given by
equation~(\ref{eq:distp_IC_nocosmo}).
For this electron spectrum, the IC radiation spectrum at high
frequencies, $\nu \gg 4 \gamma_{max}^2 k T_{CMB} / h $,
is given by:
\begin{eqnarray}
L_\nu & \approx & \frac{24 \pi h \sigma_T}{c^2} \, \Gamma ( p_0 - 1 ) \,
\left( \frac{k T_{CMB}}{h} \right)^3 \,
N_1 \left( 2 \gamma_{max} \right)^{- p_0 + 1} \nonumber \\
& & \quad \times
\left( \frac{h \nu}{4 \gamma_{max}^2 k T_{CMB}} \right)^{-p_0 + 2}
\exp \left( - \, \frac{h \nu}{4 \gamma_{max}^2 k T_{CMB}} \right) \, .
\label{eq:ic_cutoff}
\end{eqnarray}
where $\Gamma$ is the gamma function.
Thus, the spectrum falls off exponentially or slightly faster at high
frequencies.

At low frequencies, the IC spectra of these initial population only models 
flatten and become slowly increasing with frequency as the models
age.
We expect that the low energy electron spectrum in an older model with
only an initial electron population will become nearly independent of
$\gamma$ (eq.~\ref{eq:low_energy_soln} and Figure~\ref{fig:model_init_z}),
except for a slowly varying logarithmic factor.
If $N( \gamma ) \approx N_{low}$ at low energies, where $N_{low}$ is a
constant, then the IC spectrum is given by
\begin{equation} \label{eq:ic_low}
L_\nu \approx \frac{22 \pi^{3/2} h \sigma_T}{5 c^2} \, \zeta ( 5/2 ) \,
\left( \frac{k T_{CMB}}{h} \right)^{5/2} \,
N_{low} \nu^{1/2} \, ,
\end{equation}
where $\zeta ( 5/ 2 ) \approx 1.3415$ is the Riemann zeta function.
Thus, the low frequency IC spectrum varies as $L_\nu \propto \nu^{1/2}$
for older clusters with initial electron populations only.

\centerline{\null}
\vskip2.55truein
\includegraphics{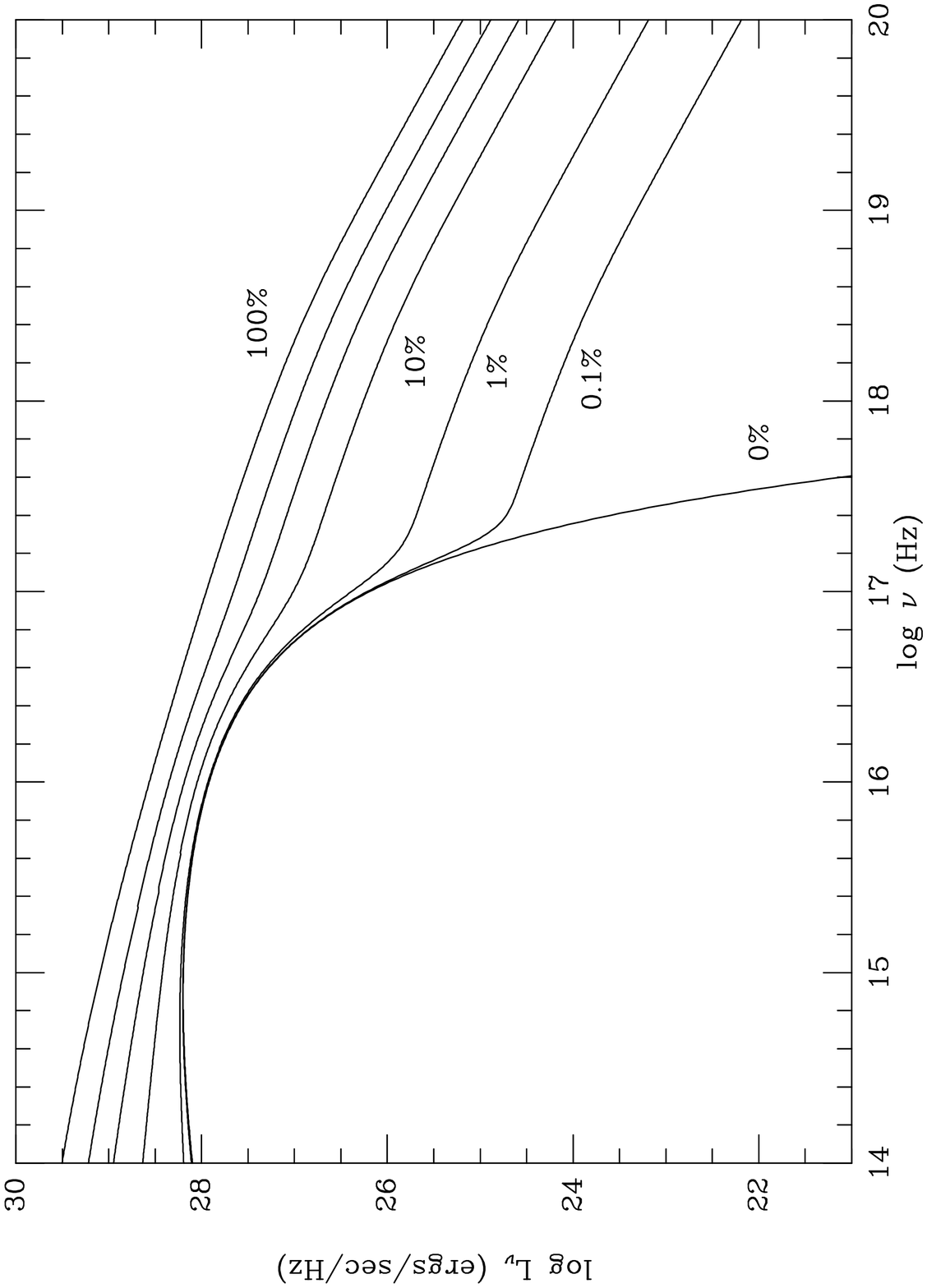}
\figcaption{
The IC emission spectra for cluster models with both an initial population
of electrons (at $z_i = 0.3$) and subsequent steady injection
(since $z_s = 0.05$).
The models are the same as those shown in Figure~\protect\ref{fig:model_both}.
The values of the fraction of the particle energy which is due to the
current injection are (top to bottom)
$F_{inj} = 100$\%, 50\%, 25\%, 10\%, 1\%, 0.1\%, and 0\%.
These are Models 23--28 and 11 in Table~\protect\ref{tab:models}.
\label{fig:ic_both}}

\vskip0.2truein

The spectra of models with both an initial population of particles
and particle injection at the present time are shown in
Figure~\ref{fig:ic_both}.
These are the models whose electron energy spectra are given in
Figure~\ref{fig:model_both} (Models 23--28 and 11).
The different curves represent models having differing fractions $F_{inj}$
of the total electron energy due to the present rate of particle injection.
The spectra in Figure~\ref{fig:ic_both} correspond to values of
$F_{inj} = 100$\%, 50\%, 25\%, 10\%, 1\%, 0.1\%, and 0\%.
Obviously, the spectrum of the models with either a pure initial
electron population ($F_{inj} = 0$\%) or no initial electron population
($F_{inj} = 100$\%) are like those given in Figures~\ref{fig:ic_initial}
and \ref{fig:ic_steady}, respectively.
Models in which the current rate of particle injection provides a
small but significant fraction of the total electron energy have
hybrid spectra.
At low frequencies ($\nu \la 10^{17}$ Hz), they have an extended hump of
emission, with a rapid fall off above $\nu \sim 10^{16}$ Hz.
However, they also have an extended hard tail of emission at high
frequencies, which has a power-law spectrum with a spectral index of
$\alpha \approx - ( p / 2 ) \approx -1.15$.

\centerline{\null}
\vskip2.55truein
\includegraphics{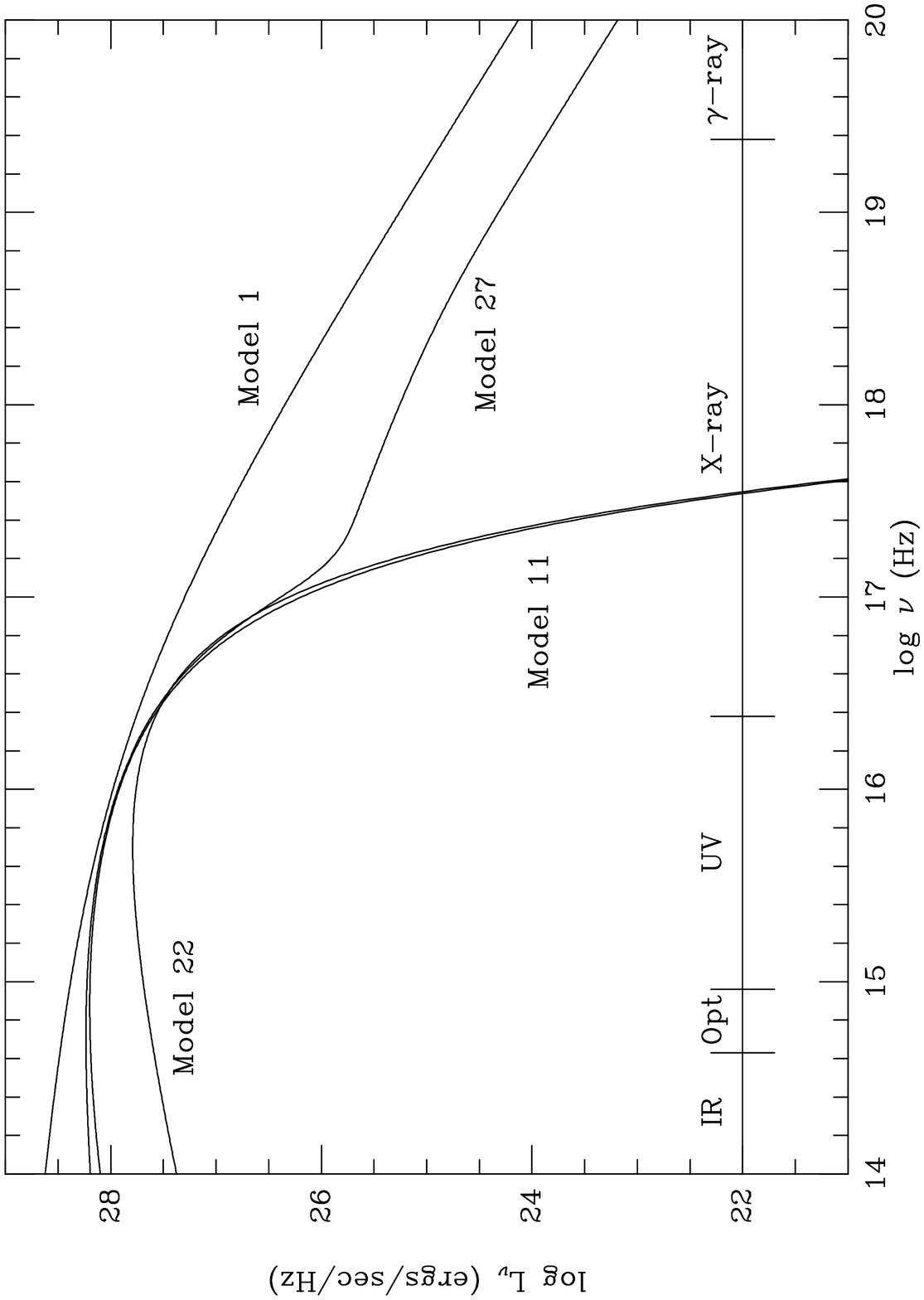}
\figcaption{
The IC emission spectra for different types of cluster models.
Models 1 has continual particle injection but no initial electron
population (Table~\protect\ref{tab:models}).
Model 11 has an initial electron population, but no subsequent electron
injection.
Model 22 is identical to Model 11, but the initial electron energy
spectrum has a break at an energy of 1 GeV.
Model 27 has both an initial electron population, and a source of
electrons at present.
The scale near the bottom of the Figure shows the portion of the
electromagnetic spectrum.
\label{fig:ic}}

\vskip0.2truein

The results for different types of models for the cluster electron
population are summarized in Figure~\ref{fig:ic}.
Models 1 has continual particle injection in steady state with
losses.
The initial electron population in Model 11 was created at $z_i = 0.3$,
but there has been no subsequent particle injection.
Model 22 is identical, but the initial electron energy spectrum had
a break at an energy of 1 GeV.
This causes the IC spectrum to decline more rapidly toward low frequencies.
Finally, Model 27 has both an initial electron population (formed at
$z_i = 0.3$) and particle injection at present (since $z_i = 0.05$) 
which contributes 1\% of the total electron population over the history
of the cluster.
At the bottom of Figure~\ref{fig:ic}, there is a scale which shows
the portion of the electromagnetic spectrum involved.
The boundaries were taken to be:
IR--Opt, 7000 \AA;
Opt--UV, 3300 \AA;
UV--X-ray, 0.1 keV;
X-ray--$\gamma$-ray, 100 keV.
We discuss several of these spectral bands separately below.
Because many other processes contribute to the gamma-ray emission by
these particles, I will discuss their observable gamma-ray properties
in another paper.
Some of the properties of the IC emission from the models for clusters
are listed in Table~\ref{tab:IC}.
The total luminosity $L_{IC}$ of the IC emission is given in the second
column.
The other entries are discussed below.

% \placetable{tab:IC}

%
% table of model ic properties
%
% SUBMISSION
% \begin{table}[thb]
% \small
% \caption{\hfil Inverse Compton Emission from Models \label{tab:IC} \hfil}
% PREPRINT
\begin{table*}[htb]
\footnotesize
\tabcaption{\hfil Inverse Compton and Radio Emission from Models
\label{tab:IC} \hfil}
\begin{center}
\begin{tabular}{lccccccccccc}
\tableline
\tableline
&$L_{IC}$&$L_\nu$ (EUV)&$L_{EUV}$&$\alpha_{EUV}$&$L_\nu$ (HXR)&$L_{HXR}$
&$\alpha_{HXR}$&$L_\nu$ (radio)&$\alpha_{radio}$&$L_\nu$ (UV)&
$\alpha_{Opt,UV}$\cr
Model&($10^{44}$ ergs s$^{-1}$)&($10^{27}$ ergs)&($10^{44}$ ergs s$^{-1}$)&&
($10^{25}$ ergs)&($10^{44}$ ergs s$^{-1}$)&&($10^{33}$ ergs)&&
$10^{28}$ ergs)&\cr
\tableline
 1&   \phn18.30&\phn6.406&1.9685&\phn$-$0.68&\phn1.82&\phn2.90&$-$1.12&\phn\phn3.41&$-$1.14&\phn2.10&$-$0.28\cr
 2&   \phn22.86&\phn8.002&2.4590&\phn$-$0.68&\phn2.27&\phn3.63&$-$1.12&\phn\phn4.26&$-$1.14&\phn2.63&$-$0.28\cr
 3&   \phn32.42&   11.352&3.4884&\phn$-$0.68&\phn3.23&\phn5.14&$-$1.12&\phn\phn6.06&$-$1.14&\phn3.70&$-$0.28\cr
 4&   \phn44.90&   15.202&4.7490&\phn$-$0.64&\phn4.52&\phn7.20&$-$1.12&\phn\phn8.49&$-$1.14&\phn4.78&$-$0.30\cr
 5&   \phn93.30&   21.628&6.9003&\phn$-$0.60&   11.04&   17.59&$-$1.12&   \phn20.75&$-$1.14&\phn7.91&$-$0.39\cr
 6&      268.00&   27.269&8.5045&\phn$-$0.64&   44.94&   74.22&$-$0.72&      172.59&$-$1.08&   13.63&$-$0.59\cr
 7&   \phn23.98&   10.906&3.0394&\phn$-$0.89&\phn1.84&\phn2.94&$-$1.12&\phn\phn3.42&$-$1.14&\phn7.30&$-$0.54\cr
 8&\phn\phn8.02&\phn4.166&1.1759&\phn$-$0.86&\phn0.58&\phn0.92&$-$1.15&      440.03&$-$1.12&\phn1.90&$-$0.33\cr
 9&      209.07&   26.360&8.2593&\phn$-$0.63&   38.87&   62.39&$-$0.85&   \phn46.61&$-$3.62&   12.31&$-$0.53\cr
10&   \phn26.44&   16.356&5.2954&\phn$-$0.57&\phn0.00&\phn0.00&  ---  &\phn\phn0.00&   --- &\phn4.25&$-$0.14\cr
11&\phn\phn2.86&\phn4.788&1.0884&\phn$-$1.47&\phn0.00&\phn0.00&  ---  &\phn\phn0.00&   --- &\phn1.52&  +0.00\cr
12&\phn\phn0.18&\phn0.004&0.0005&   $-$11.93&\phn0.00&\phn0.00&  ---  &\phn\phn0.00&   --- &\phn0.41&$-$0.29\cr
13&\phn\phn0.00&\phn0.000&0.0000&     ---   &\phn0.00&\phn0.00&  ---  &\phn\phn0.00&   --- &\phn0.00& ---   \cr
14&   \phn11.59&   14.505&3.4343&\phn$-$1.33&\phn0.00&\phn0.00&  ---  &\phn\phn0.00&   --- &\phn9.25&$-$0.45\cr
15&\phn\phn8.48&   10.553&3.0849&\phn$-$0.80&\phn0.00&\phn0.00&  ---  &\phn\phn0.00&   --- &\phn2.65&$-$0.06\cr
16&\phn\phn2.89&\phn4.828&1.1023&\phn$-$1.45&\phn0.00&\phn0.00&  ---  &\phn\phn0.00&   --- &\phn1.52&  +0.00\cr
17&\phn\phn2.63&\phn4.393&0.9561&\phn$-$1.60&\phn0.00&\phn0.00&  ---  &\phn\phn0.00&   --- &\phn1.51&$-$0.01\cr
18&\phn\phn1.31&\phn1.510&0.2235&\phn$-$3.31&\phn0.00&\phn0.00&  ---  &\phn\phn0.00&   --- &\phn1.37&$-$0.07\cr
19&\phn\phn0.40&\phn0.013&0.0016&   $-$11.03&\phn0.00&\phn0.00&  ---  &\phn\phn0.00&   --- &\phn0.87&$-$0.29\cr
20&   \phn13.18&\phn4.826&1.4753&\phn$-$0.69&\phn1.27&\phn2.03&$-$1.12&\phn\phn2.37&$-$1.14&\phn1.58&$-$0.27\cr
21&   \phn18.32&\phn2.674&0.9804&\phn$-$0.32&\phn2.45&\phn3.90&$-$1.12&\phn\phn4.59&$-$1.14&\phn0.34&  +0.10\cr
22&\phn\phn2.23&\phn4.290&1.1006&\phn$-$1.16&\phn0.00&\phn0.00&  ---  &\phn\phn0.00&   --- &\phn0.54&  +0.27\cr
23&      140.48&   24.309&7.6864&\phn$-$0.61&   20.77&   32.97&$-$1.10&   \phn39.17&$-$1.14&   10.12&$-$0.46\cr
24&   \phn71.65&   14.548&4.3874&\phn$-$0.71&   10.39&   16.48&$-$1.10&   \phn19.58&$-$1.14&\phn5.75&$-$0.42\cr
25&   \phn37.26&\phn9.668&2.7379&\phn$-$0.86&\phn5.19&\phn8.24&$-$1.10&\phn\phn9.79&$-$1.14&\phn3.64&$-$0.35\cr
26&   \phn16.62&\phn6.740&1.7482&\phn$-$1.07&\phn2.08&\phn3.30&$-$1.10&\phn\phn3.92&$-$1.14&\phn2.37&$-$0.23\cr
27&\phn\phn4.23&\phn4.983&1.1544&\phn$-$1.40&\phn0.21&\phn0.33&$-$1.10&\phn\phn0.39&$-$1.14&\phn1.61&$-$0.04\cr
28&\phn\phn3.00&\phn4.800&1.0945&\phn$-$1.46&\phn0.02&\phn0.030&$-$1.10&\phn\phn0.04&$-$1.14&\phn1.53&$-$0.01\cr
\tableline
\end{tabular}
\end{center}
% SUBMISSION
% \end{table}
% PREPRINT
\end{table*}

\subsection{EUV and Soft X-ray Emission} \label{sec:ic_euv}

Figure~\ref{fig:ic} shows that the fluxes from different models all tend
to agree at frequencies $\sim$$2 \times 10^{16}$ Hz, which corresponds
to a photon energy of $\sim$80 eV.
Thus, EUV and very soft X-ray emission might be expected to be a universal
property of clusters of galaxies, if they have all produced populations
of relativistic electrons with total energies which are at least
$\sim$$10^{-2}$ of their thermal energies.
The one constraint is that at least a portion of these particles
must have been injected at moderate to low redshifts, $z_i \la 1$.
The fact that essentially all of the models produce a strong and
similar flux at EUV energies may help to explain the fact the excess
EUV emission has been detected in all of the cluster observed with the
{\it Extreme Ultraviolet Explorer} ({\it EUVE})
satellite and which lie in directions of sufficiently low Galactic
columns that this radiation is observable
(Lieu et al.\ 1996a,b;
Mittaz, Lieu, \& Lockman 1998;
Sarazin \& Lieu 1998).
Values of $L_\nu (EUV)$ at a frequency of $v = 2 \times 10^{16}$ Hz
are listed in column 3 of in Table~\ref{tab:IC}.
From these values, one finds that the emission at EUV energies is
fairly directly related to 
$E^{tot}_{CR,e}$, the total amount of energy injected in electrons
with $\gamma \ge 300$.
The average relationship is about
\begin{equation} \label{eq:euv_lum}
L_\nu (EUV) \sim 6 \times 10^{27} \,
\left( \frac{E^{tot}_{CR,e}}{10^{63} \, \rm{ergs}} \right) \, {\rm ergs}
\, .
\end{equation}
Models which are likely to apply to real clusters have spectrum which
agree with equation~(\ref{eq:euv_lum}) to within a factor of 4.
I have also calculated the total luminosity in the {\it EUVE} observing band of
photon energies of 65 to 245 eV.
These values of $L_{EUV}$ are given in column 4 of Table~\ref{tab:IC}.

While the amount of EUV emission is fairly constant from model to model,
the spectrum depends strongly on the amount of recent particle injection.
I have fit the model spectra in the {\it EUVE} band of 65 to 245 eV to a
power-law spectrum of the form $L_\nu \propto \nu^\alpha$.
Specifically, $\log L_\nu$ was fit to a linear function of $\log \nu$ using the
least squares method.
The best-fit values of the energy spectral index $\alpha_{EUV}$ are listed
in column 5 of Table~\ref{tab:IC}.
Large negative values occur for models without current particle injection
and with small values of $\gamma_{max}$, where the IC spectrum is falling
off exponentially
(eq.~\ref{eq:ic_cutoff}).
Thus, a power-law may not be a good fit to the spectrum.
For most of the models, the spectral index lies in the range from
about -0.6 to -1.5, but very large negative values are also possible.
The spectrum drops more rapidly with frequency as the proportion of
electrons which are currently being injected decreases.
This is most evident in Models 23-28, which are identical except for a
decreasing fraction of relativistic electrons due to ongoing particle
injection.

Because the {\it EUVE} observations of clusters have no spectral resolution,
there is no detailed information on the observed spectrum of EUV 
emission
(Lieu et al.\ 1996a,b;
Mittaz, Lieu, \& Lockman 1998).
However, the ratio of {\it EUVE} fluxes to those in the softest bands of the
{\it ROSAT} PSPC suggest that the EUV spectra are generally steeply
declining.
The spectra of the Coma cluster is less steeply declining than that of
A1795;
this might be understood as the result of particle acceleration by merger
shocks, since Coma appears to be undergoing a merger or mergers
(e.g., Burns et al.\ 1994;
Donnelly et al.\ 1999),
while A1795 appears regular and relaxed
(e.g., Briel \& Henry 1996)
Also, the EUV spectra appear to get steeper with increasing radius in
the A1795 cluster
(Mittaz, Lieu, \& Lockman 1998).
This might be the result of decreasing gas density with radius;
compare Models 14 and 11 in Table~\ref{tab:IC} and
Figure~\ref{fig:model_init2}.

\subsection{Hard X-ray IC Emission and Radio Halos} \label{sec:ic_hxr}

At photon energies of 0.5 to 10 keV,
the dominant emission in most clusters is the thermal emission from the
hot ICM, and IC emission will be difficult to detect.
However, at energies $\ga$20 keV, IC emission should again become
observable, assuming that it is dropping as power-law function of
frequency, while the thermal emission drops as an exponential.
In Table~\ref{tab:IC}, the hard X-ray (HXR)
power $L_\nu$ at $\nu = 10^{19}$ Hz (41.4 keV, col.~6),  
the luminosity $L_{HXR}$ from 20 to 100 keV, and
best-fit HXR power-law spectral index $\alpha_{HXR}$
from 20 to 100 keV are given for all of the models in Table~\ref{tab:models}.
IC emission at photon energies of $\sim$50 keV will be produced by
electrons with $\gamma \sim 10^4$.
These particles have rather short lifetimes ($t_{loss} \ll 10^9$ yr),
and are only present in clusters in which there has been substantial
particle injection since $z_i < 0.1$.
As Table~\ref{tab:IC} shows, only the models with current (or very recent,
as in Model 9) particle injection have any significant HXR emission.

Because of the short lifetimes of the particles producing HXR emission,
these electrons are likely to be close to steady-state if present in
significant quantities.
The expected steady-state spectral index if IC losses dominated would be
$\alpha_{HXR} = - ( p + 1 ) / 2 \approx -1.65$.
The best-fit spectral indices are flatter than this,
$\alpha_{HXR} \approx -1.1$, mainly because other loss processes are
important at the lower energy end of the HXR band
(Fig.~\ref{fig:losses}).
With the exception of the two models in which electrons have only
been injected in the cluster extremely recently (at $z_i = 0.01$, 
Models~6 and 9),
all of the HXR spectral shapes are very similar, as expected for
steady-state populations.
The differences in the HXR luminosities just reflect differences in
the present rate of particle injections.
To a good approximation, the present day value of $L_{HXR}$
(20--100 keV) is simply given by
\begin{equation} \label{eq:hxr_lum}
L_{HXR} \approx 0.17 {\dot{E}}_{CR,e} ( \gamma > 5000 ) \, .
\end{equation}
where ${\dot{E}}_{CR,e} ( \gamma > 5000 )$ is the total present rate of
injection of energy in cosmic ray electrons with $\gamma > 5000$.
The best-fit coefficient (0.17 in eqn.~\ref{eq:hxr_lum}) depends somewhat
on the power-law index of the injected electrons; the value of 0.17 applies
for $p = 2.3$.

\centerline{\null}
\vskip2.55truein
\includegraphics{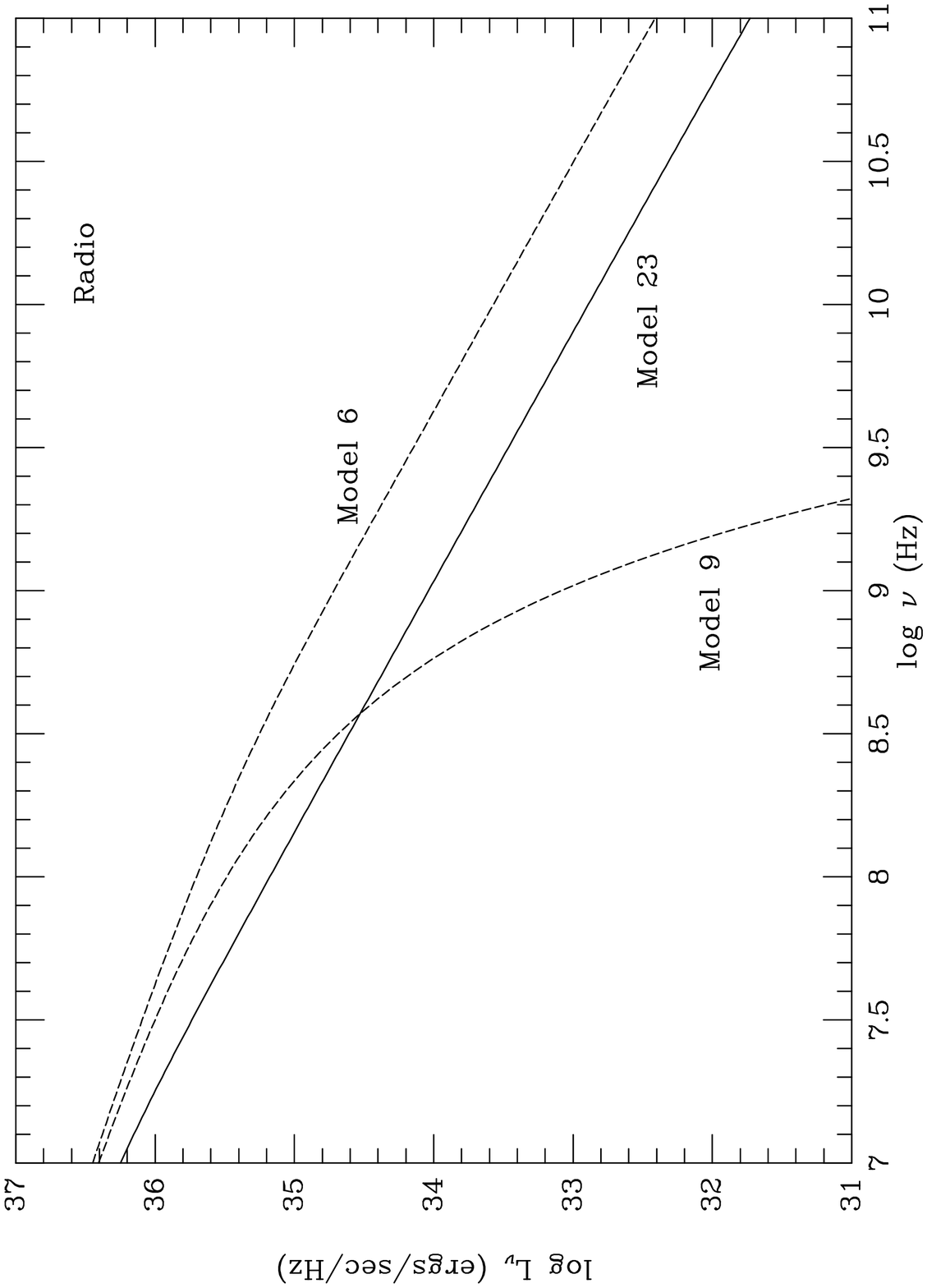}
\figcaption{
The radio synchrotron emission from three cluster models.
The solid curve gives the radio spectrum of Model 23, which is
typical of the models with current particle injection which has
reached steady-state.
Almost all of the models which have significant radio emission have
spectra which are nearly identical to Model 23.
In Model~6, the electrons haven't reached steady-state, and the spectrum
is somewhat flatter than that of Model 23.
On the other hand, in Model 9, all of the electrons were injected at
$z_i = 0.01$, and the highest energy electrons have lost energy by
IC and synchrotron emission.
This spectrum is much steeper than Model 23 at high frequencies.
\label{fig:radio}}

\vskip0.2truein

The same relativistic electrons which produce HXR emission by IC scattering
will also produce radio emission by synchrotron emission.
We have calculated the synchrotron radio emission from our models for
the intracluster electron population.
Averaging over all orientations of the magnetic field and all pitch
angles for the electrons gives a radio power of
\begin{equation} \label{eq:syn_power}
L_\nu = \frac{{\sqrt{3}} e^3 B}{m_e c^2} \int N( \gamma ) R[ x ( \gamma ) ]
\, d \gamma \, .
\end{equation}
The function $R(x)$ is defined as
(Ghisellini, Guilbert, \& Svensson 1988)
\begin{equation} \label{eq:syn_r}
R ( x ) \equiv 2 x^2 \left\{
K_{4/3} (x) K_{1/3} (x) - \frac{3}{5} x \left[
K_{4/3}^2 (x) - K_{1/3}^2 (x) \right] \right\} \, ,
\end{equation}
where $K$ is the modified Bessel function.
The normalized frequency variable $x$ is
\begin{equation} \label{eq:syn_x}
x \equiv \frac{\nu}{3 \gamma^2 \nu_c} \, ,
\end{equation}
where $\nu_c$ is the cyclotron frequency
\begin{equation} \label{eq:cyclo}
\nu_c \equiv \frac{e B}{2 \pi m_e c} \, .
\end{equation}
The values of the magnetic fields in the models are given
in Table~\ref{tab:models}.
For an electron with a given energy, the radio spectrum in
equation~(\ref{eq:syn_power}) is maximum for $x \approx 0.1146$.
Electrons with a Lorentz factor of $\gamma$ produce radio emission
with $\nu \sim 100 ( B / \mu G) ( \gamma / 10^4 )^2 $ MHz, and rather
high electron energies ($\gamma \ga 10^4$) are needed to produce
observable radio emission.
Thus, these are likely to be the same electrons which produce the hardest
HXR emission.

Column~9 of Table~\ref{tab:IC} gives the resulting radio powers of the
models at a wavelength of 91 cm ($\nu = 328$ MHz), while column~10
gives the spectral index between 91 and 21 cm ($\nu = 1.4$ GHz).
The results are very similar to those for the HXR emission.
Because of the short lifetimes of the high energy electrons required
for radio emission,
only models with very recent particle injection (at $z \la 0.01$) have
any significant radio emission.
The high energy particles tend to be in steady state, and the radio
spectra of the models tend to be quite similar.
Three of the most divergent radio spectra are shown in Figure~\ref{fig:radio}.
The radio spectrum of Model 23 is typical of those for steady-state
models with continual particle injection, and applies to most of the models
with significant radio emission.
In Model~6, the electrons have not yet reached steady-state,
and the radio spectrum is flatter.
All of the electrons in Model~9 were injected at $z_i = 0.01$, and
the highest energy particles have been removed by IC and synchrotron
losses.
This spectrum is quite steep.

In general, HXR emission and radio synchrotron emission are expected only
in clusters with very recent or current injection of relativistic
electrons.
Thus, both measure the current rate of particle injection.
For example, if the particles are accelerated in ICM shocks, HXR and
radio emission would be expected only in clusters which are currently
undergoing (or which very recently underwent) a merger.

\subsection{Optical and UV Emission} \label{sec:ic_optuv}

One also expects that the lower energy portion of the cosmic ray population
in clusters will produce diffuse optical and UV emission.
Diffuse optical emission is known to exist in many clusters, particularly
in those with central cD galaxies
(e.g., Boughn \& Uson 1997).
Although the origin is not completely understood, the optical colors
of the diffuse light suggest that it is due to old stars, which may have
been stripped from cluster galaxies.
It is likely that the near or vacuum UV are better regions to detect
low surface brightness diffuse emission due to IC emission,
since the older stellar population in E and S0 galaxies (and, presumably,
the intracluster stellar population) are fainter there.

Table~\ref{tab:IC} gives the power $L_\nu (UV)$ at a wavelength of
2000 \AA\ and the best-fit power-law spectral index $\alpha_{Opt,UV}$
between 2000 \AA\ and 8000 \AA.
The UV powers of most of the models lie within a relatively narrow range
of about an order of magnitude.
Unless the electron density in the ICM is much higher than
0.001 cm$^{-3}$ and Coulomb losses are catastrophic or the electron
population in clusters is very old, one would always expect a significant
population of lower energy electrons.
The IC spectra in the optical and UV are fairly flat, with spectral indices
of $-0.5 \ga \alpha_{Opt,UV} \ga 0.3$.
In models with current particle injection, the lower energy electron
population should be approaching steady-state with the Coulomb
losses (eq.~\ref{eq:steady_low}).
Then, the expected power-law for the IC emission is
$\alpha \approx - ( p - 2 ) / 2 \approx -0.15$.
For models with only an initial electron population, the electron
spectrum is expected to be nearly flat at low energies, and the
spectrum is given by equation~(\ref{eq:ic_low}).
Thus, the spectral index should be slightly positive in these models.

The predicted fluxes are rather low when compared to the diffuse optical
emission seen in clusters of galaxies or to the sensitivity of current
and UV instruments for detecting diffuse emission.

\section{Conclusions} \label{sec:conclude}

Models for the integrated population of primary cosmic ray electrons in
clusters of galaxies have been calculated.
The evolution of the relativistic electrons included the effects of energy
losses due to IC scattering, synchrotron emission, Coulomb losses to the
ICM, and bremsstrahlung.
For typical cluster parameters, the combined time scale for these losses
reaches a maximum of $\sim 3 \times 10^9$ yr for electrons with a
Lorentz factor $\gamma \sim 300$.
This maximum loss time scale is comparable to the Hubble time or the typical
age of clusters.
For relativistic electrons with either much higher or lower energies, the
loss time scale is considerably shorter than the typical age of a cluster.

Although the models don't depend on the detailed nature of the
source of the relativistic electrons, we assume that they are
accelerated by shocks, either due to the formation of the cluster
and subcluster mergers, or due to radio galaxies.
We present models in which the electrons are all injected at some
given time in the past (e.g., the epoch of cluster formation or the
time when a powerful radio galaxy is active).
We also consider models with continuous electron injection, either
at a constant rate or at a rate given by the self-similar solution
for secondary infall onto clusters
(Bertschinger 1985).
Hybrid models with both initial electron populations and continual
particle injection were also calculated.
Based on the results for particle acceleration is supernova shocks,
in most of the models we assume that the energy spectrum of the injected
electrons is a power-law with an exponent of 2.3.
We also consider a few models with a broken power-law, where the break
occurs at the energy at which protons become relativistic.

Simple analytical solutions are given for models in which the 
electrons are either all injected at a single time in the past, or in
which there is a constant rate of injection of electrons but no initial
electron population.
These solutions apply in the limits of either high or low energies, where
IC and synchrotron or Coulomb losses dominate.
For models in which all of the electrons were injected in the past,
there is a high energy cutoff to the present electron distribution
for $\gamma \ge \gamma_{max}$.
Expressions for the resulting populations and $\gamma_{max}$ were given
which included synchrotron losses and the cosmological variation in IC
losses due to the redshift of the CMB radiation field.
At low energies where Coulomb losses dominate, the electron distribution
function $N( \gamma )$ tends to a constant value, independent of $\gamma$.

Analytic solutions were also given for models with constant electron
injection.
At very low or high energies, the particle distribution approaches steady
state.
If the electrons are injected with a power-law distribution, the steady
state distribution is one power steeper at high energies and one
power flatter at low energies
(e.g., Ginzburg \& Syrovatskii 1964).
A self-similar solution which connects these steady-state solutions to
the time-dependent solution when the losses are not rapid was also derived.

The numerical models for the electron population are normalized such that
the total energy of relativistic electrons injected with $\gamma \ge 300$
is $10^{63}$ ergs.
For most of the models, this leads to a present-day energy in electrons
with $\gamma \ge 300$ of $\sim$$10^{62}$ ergs.
Because the equations for the electron population are linear, models for
other values can be derived by simple scaling from these values.

The most important factor affecting the energy distribution of cosmic ray 
electrons in the cluster models is the history of injection of the
electrons.
The numerical models show that only clusters in which there has been a
substantial injection of relativistic electrons since $z \la 1$ will
have any significant population of primary cosmic ray electrons at
present.
For older electron populations, the high energy electrons are removed by
IC losses, whose rate increases with redshift.
Lower energy electrons (including initially higher energy particles
which have lost energy by IC scattering) are removed by Coulomb
losses.

In models with current (or very recent, $z \ll 0.1$) particle injection,
the electron population is large at low energies and extends to very
high energies.
In models with ongoing electron injection, the distribution quickly
reaches steady-state at high ($\gamma \ga 300$)
and low ($\gamma \la 300$) energies.
Due to IC losses, the high energy electron distribution is steeper than the
injected one (by one exponent for a power-law injection spectrum).
Coulomb losses cause the low energy distribution to be flatter than the
injected spectrum (again, by one exponent for a power-law injection).

In models in which there is no current or recent particle injection, the
high energy electron distribution has a cutoff for
$\gamma \ge \gamma_{max}$.
Thus, no very high energy primary electrons should be present.
The low energy distribution is also considerably flatter than that in
models with current particle injection.
In models with a large initial population of particles, but also with
a significant rate of current particle injection, the electron
distributions are a simple combination the behavior of the initial
population models and the steady injection models.
There is a steep drop in the electron population at $\gamma_{max}$,
but higher energy electrons are present at a rate determined by the
current rate of particle injection.
The electron population at $\gamma \la \gamma_{max}$ is fairly
flat, but eventually start to increase significantly with decreasing
energy at a lower energy.

Increasing (decreasing) the ICM thermal gas density decreases
(increases) the number of low energy electrons ($\gamma \la 100$),
since these electrons are removed by Coulomb losses to the ambient plasma.
If the magnetic field is greater than generally expected over most of the
volume of a cluster ($B \ga 3$ $\mu$G), synchrotron losses will reduce the
number of high energy electrons beyond the effect of IC emission.

The portion of the electron spectrum where the different models exhibit
the greatest level of agreement is the region near $\gamma \sim 300$,
where the loss time of the electrons is maximum and where it reaches values
of $\sim 3 \times 10^9$ yr, which are comparable to cluster ages.
The general prediction is at all clusters should contain a significant
electron population at these energies, as long as there has been
significant particle injection since $z \la 1$.

I also calculated the IC and synchrotron emission from these models.
In models with steady particle injection with a power-law exponent $p$,
the IC spectra relax into a steady-state form.
At low energies, the spectrum is a power-law with 
$\alpha \approx - ( p - 2 ) / 2 \approx -0.15$, while at high energies
it is a power-law with $\alpha \approx - ( p / 2 ) \approx -1.15$.
These two power-laws meet at a knee at $\nu \sim 3 \times 10^{16}$ Hz.
In models with no current particle injection, the cutoff in the electron
distribution at high energies ($\gamma \ge \gamma_{max})$ results in
a rapid drop in the IC spectrum at high frequencies.
At low frequencies, the IC spectra of these models are fairly flat or even
slightly rising with frequency.
The IC spectra of hybrid models with both an initial population of particles
and particle injection at the present time are intermediate.
In models in which the current rate of particle injection provides a
small but significant fraction of the total electron energy,
the spectra show an extended hump at low frequencies ($\nu \la 10^{17}$ Hz),
with a rapid fall off above $\nu \sim 10^{16}$ Hz.
However, they also have an extended hard tail of emission at high
frequencies, which has a power-law spectrum with a spectral index of
$\alpha \approx - ( p / 2 ) \approx -1.15$.

In the models, EUV and soft X-ray emission are nearly ubiquitous.
This emission is produced by electrons with $\gamma \sim 300$, which
have the longest lifetimes in clusters.
The lifetimes of these particles approach the likely ages of clusters,
and thus particles at this energy are likely to be present in most
clusters.
This may explain why {\it EUVE} has detected
EUV emission from all of the cluster observed which lie in directions of
sufficiently low Galactic columns
(Lieu et al.\ 1996a,b;
Mittaz, Lieu, \& Lockman 1998;
Sarazin \& Lieu 1998).
In the models, the increase in the IC losses in going from the EUV to the
soft X-ray region causes the spectral to decline rapidly in this region.
More steeply declining spectra are expected in clusters without current
sources of particles or with lower thermal gas densities.
Assuming that the acceleration of relativistic electrons in clusters is
due to ICM shocks, this may explain why the EUV spectrum in the very
regular cluster A1795 is steeper than that in Coma 
(Mittaz, Lieu, \& Lockman 1998), and why the spectrum in A1795 gets steeper
at larger radii.

The IC emission also extends down in frequency into the UV, optical, and IR
bands.
The spectrum in this region is expected to be fairly flat, with power-law
spectral indices in the range from -0.6 to +0.3.
The steeper spectra are for models with significant current particle
acceleration, while the flatter or rising spectra are for models with
older electron populations and/or broken power-law electron acceleration.
In the optical, the luminosities and expected surface brightnesses are below
the level of diffuse stellar light observed in clusters.
This emission might be easier to detect in the vacuum UV, where the older
stars associated with E/S0 galaxies are fainter.
However, the surface brightnesses are rather low.
The UV and EUV emission from clusters might make a contribution
to the diffuse extragalactic ionizing radiation, although crude estimates
suggest this will not be very significant.

In much of the X-ray spectral band (from 0.5 to 10 keV),
the dominant luminosity in most clusters is the thermal emission from the
hot ICM, and IC emission will be difficult to detect.
However, at energies $\ga$20 keV, IC emission can again become
observable.
I calculated the hard X-ray (HXR) IC emission from the cluster models.
Radiation at these energies is produced by high energy electrons
($\gamma \sim 10^4$) which have rather short lifetimes.
As a result, this emission is only expected in clusters which have
current (or very recent) particle acceleration.
The luminosity of HXR emission is primarily determined by the current
rate of particle acceleration in the cluster.
The same particles which produce HXR radiation by IC scattering will
also produce radio emission by synchrotron emission at observable
frequencies.
I have also calculated the radio spectra produced by the models.
In general, the high energy electrons which produce the HXR and radio
emission should be nearly in steady-state, and the predicted spectra for
most models with significant HXR or radio emission are approximately
power-laws with spectral indices of $\alpha \approx -1.1$.

If the main source of primary relativistic electrons is acceleration
in ICM shocks, then HXR tails and diffuse radio halos should be confined
to clusters which currently have strong ICM shocks.
Thus, we would expect that these would mainly be clusters presently
undergoing significant mergers.
In the case of radio halos, there is some evidence that this is indeed the
case
(e.g., Tribble 1993).
At the same time, large cluster radio halos are quite rare.
This may be related to the short lifetimes of the high energy particles
required and to the need for the merger to be at a stage where the shocks
are very strong;
alternatively, some other physical properties
(e.g., magnetic field intensity or geometry, or high shock Mach number, \dots)
may be required to produce high energy electrons in ICM shocks.

Because the HXR and radio halo emissions are produced by particles with
very short lifetimes while the EUV emission is due to electrons
whose lifetimes are a significant fraction of the likely ages of clusters,
one would not expect the same power-law spectrum which fits the HXR and
radio emission to extend to the EUV emission.
In fact, EUV emission is seen from clusters with no detectable radio halo
emission (such as A1795), so it is clear that the EUV and HXR/radio cannot
be connected with a single power-law spectrum with a normal spectral
index.
The EUV and HXR/radio should connect continuously in clusters where a large
fraction of the cosmic ray electrons are being generated at the present
time (i.e., in an ongoing or very recent strong merger).
In the more general case, one expects a broad peak in the IC emission
in the EUV band, a rapid drop-off in the spectrum from the EUV to the
X-ray band, and HXR/radio emission proportional to the present rate
of particle acceleration (Figures~\ref{fig:ic_both} and \ref{fig:ic}).

\acknowledgements
This work was supported in part by NASA Astrophysical Theory Program grant
NAG 5-3057.
I thank Richard Lieu and Andy Fabian for helpful conversations or
correspondence.
I would also like to thank the referee, Bob Gould, for a very helpful
and expeditious report.

\end{document}